\documentclass[12pt]{article}
\date{\today}

\def\be{\begin{equation}}
\def\ee{\end{equation}}
\def\bear{\begin{eqnarray}}
\def\eear{\end{eqnarray}}

\def\half{{ \frac{1}{2} }}

\def\p{{\partial}}

\def\dg{{\dagger}}

\def\Id{{\bf I}} 
\def\Z{{\bf Z}} 
\def\barn{{\overline n }}    
\def\bart{{\overline t }}  
\def\barh{{\overline h}}
\def\bary{{\overline y }}
\def\barx{{\overline x }}  
\def\barE{{\overline E }}     
\def\barG{{\overline G }}   
\def\barD{{\overline D }}
\def\barA{{\overline A }}
\def\barL{{\overline L }}
     
\def\barm{{\overline m }}  
\def\barg{{\overline g }}    
\def\bars{{\overline s }}  
\def\barf{{\overline f }}
\def\bargamma{{\overline \gamma }}    
\def\barnu{{\overline \nu }}                                     
    
\def\bG{{\bf G}}  
\def\bB{{\bf B}} 
\def\bC{{\bf C}}  
\def\barbB{\overline{{\bf B }}} 
\def\barbC{\overline{ {\bf C}} }   
\def\bH{{\bf H}}
\def\bA{{\bf A}}  
\def\barbA{{\overline {\bf  A }}}                                    
\def\wt{\widetilde}
\def\tbG{\wt{{\bf G}}}     
                            
\def\a{{\alpha}}
\def\b{{\beta}}
\def\c{{\gamma}}
\def\d{{\delta}}

\def\h{{\eta}}

\def\tht{{\theta}}
\def\btht{\overline{{\theta}}}
\def\barmu{\overline{{\mu}}}

\def\bara{{\overline{\alpha}}}
\def\barc{{\overline{\gamma}}}
\def\barb{{\overline{\beta}}}
\def\bard{{\overline{\delta}}}

\def\bpsi{{\overline{\psi}}}
\def\barz{{\overline{z}}}

\def\al{{{\mbox{\boldmath $\scriptstyle \a$} }}}
\def\bel{{{\mbox{\boldmath $\scriptstyle \b$} }}}
\def\cel{{{\mbox{\boldmath $\scriptstyle \c$} }}}
\def\del{{{\mbox{\boldmath $\scriptstyle \d $} }}}
\def\albar{{{\mbox{\boldmath $\scriptstyle \bara$} }}}
\def\belbar{{{\mbox{\boldmath $\scriptstyle \barb$} }}}
\def\celbar{{{\mbox{\boldmath $\scriptstyle \barc$} }}}
\def\delbar{{ {\mbox{\boldmath $\scriptstyle \bard $}} }}
\def\mw{{ {\mbox{$w$} }}}
\def\my{{ {\mbox{$y$} }}}
\def\mf{{ {\mbox{$f$} }}}
\def\barmw{{  {\displaystyle{ \bar w } } }}
\def\barmy{{  {\displaystyle{ \bar y } } }}
\def\barmf{{  {\displaystyle{ \bar f } } }}
\def\baru{{\overline{u}}}
\def\barl{{\overline{l}}}
\def\barD{{\overline{D}}}
\def\barcalD{{\overline{{\cal D}}}}
\def\e{{\epsilon}}
\def\l{{\tilde l}}
\def\h{{\tilde h}}

\def\hC{{\hat C}}
\def\hA{{\hat A}}
\def\hM{{\hat M}}
\def\hN{{\hat N}}
\def\hP{{\hat P}}

\def\hF{{\hat F}}
\def\hD{{\hat D}}
\def\hB{{\hat B}}

\def\A{{\tilde A}}

\def\G{{\tilde G}}
\def\O{{\tilde \Omega}}
\def\T{{\tilde T}}
\begin{document}
%
\begin{titlepage}
\titlepage
\rightline{hep-th/0409175}
\rightline{\today}
\vskip 3.5cm
\centerline{{\Huge  Geometric interpretation of the large $N=4$ index }}
\vskip 2cm
\centerline{  Natalia Saulina  }
\vskip 1cm
\begin{center}
 Jefferson Physical Laboratory
 \\  Harvard University \\
Cambridge, MA 02138, USA\\
\end{center}
\vskip 2cm

\abstract{ We study WZW models with the large $N=4$ superconformal symetry.
Our main result  is a geometric interpretation of the large $N=4$ index. 
In particular, we find that states contributing to the index
belong to  spectral flow orbits of special RR ground states.
We use  (anti-)holomorphic  differentials
with torsion to clarify the geometric meaning
of these states in terms of  differential forms 
on the target space. 

}
\end{titlepage}

\section{Introduction}
 Supersymmetric 2d quantum field theories have been an area of active
research for several decades.
Supersymmetry is a powerful tool
which gives a lot of control over the structure of these theories.
In particular,  it   allows
the construction of indices \cite{Wzero,SW,W,AKMW,CFIV} which
provide important
information about  the theory at a generic point in the moduli space. 
For example, the elliptic genus 
\cite{ SW, W, AKMW}  was used to establish
the relation between $N=2$ minimal models and Landau-Ginzburg
theories. As another example, the ``new index'' of \cite{CFIV}
was crucial for understanding the structure of  the vacua of 
the $N=2$ supersymmetric theory.

In 2d non-linear $\sigma$-models with superconformal symmetry
an index  may be understood more intuitively 
in terms of the geometry of the target space. For instance, the elliptic genus 
was interpreted in \cite{W}  as
the index of Dirac operator in loop spaces. 
The purpose of this paper is  to find   a
geometric interpretation of   the large $N=4$ index 
recently constructed in~\cite{GMMS}. 

The large $N=4$ index ($I_2$) describes
theories with the large $N=4$ superconformal symmetry  $A_{\c}$ \cite{STP}.
 The algebra $A_{\c}$ is different from  the  small
$N=4$ superconformal algebra in  that it contains
two $\widehat{su(2)}$ current algebras instead of one.
This higher symmetry  
restricts the spectrum of the theory in  essential ways \cite{GPTV,PTone,PT,OPT}.
The basic information about the spectrum, captured by the large $N=4$ index, was used
in the search \cite{GMMS,EFGT,BPS,M} for the
holographic dual CFT of the string theory on $AdS_3\times S^3\times S^3\times S^1.$ 

The  index  $I_2$  is defined by:
\be
I_2=\Biggl [z_+ {\barz}_+ \frac{d}{dz_- } \frac{d}{d{\barz}_-}Z \Biggr]_{
z_-=-z_+,\ {\barz}_-=-{\barz}_+}
\label{intro}
\ee
Here $Z$ is the RR sector $A_{\gamma}$ character \cite{PT}
\be
Z(q,z_{\pm};{\overline q},{\barz}_{\pm}):=
\mbox{Tr}_{{\cal H}_{RR}}z_-^{2iA_0^{-;3}}{\barz}_-^{2i{\overline A}_0^{-;3}}
z_+^{2iA_0^{+;3}} {\barz}_+^{2i{\overline A}_0^{+;3}}q^{L_0-c/24} {\overline q}^{{\overline L}_0-c/24}
\label{introii}
\ee 
and $A_0^{\pm;i}$ are zero modes of the $\widehat{su(2)}_+ \oplus \widehat{su(2)}_-$
currents.

The $N=4$ world-sheet supersymmetry requires \cite{SSTV,HP}
the target space $X$ of the non-linear $\sigma$-model to be a Hyper-K\" ahler-Torsion  manifold
 \cite{HOP}. However, realization of $A_{\c}$ algebra
is known only for the $N=4$ gauged  WZW models which are special due to
their relation with the $N=4$ coset CFT's \cite{SSTVtwo,GK}. For this reason,  our strategy to
reveal the geometric meaning of the large $N=4$ index $I_2 $ will rely on
these models.

The main result of this paper is a geometric interpretation  of the index $I_2$
for the $N=4$ gauged WZW models.  
We show that  the states contributing to the index $I_2$ 
belong to spectral flow orbits of special RR ground states and
characterize these states  geometrically as certain $(p_1,p_2)$ forms
on the target space $X.$ In particular, these forms satisfy
\be
{\cal D} C= \nu \wedge C,\quad \quad {\cal D}^{\dg} C=i_{\mw} C
\label{introxx}
\ee
\be
{\barcalD} C= \barnu \wedge C,\quad \quad {\barcalD}^{\dg} C=i_{ \barmw} C
\label{introxxii}
\ee
where $\left( {\barcalD} \right) \, {\cal D}$ are (anti-)holomorphic differentials
  with torsion.  Vectors $\mw,\barmw$ and 1-forms $\nu,\barnu$ 
depend on a particular RR state.

The $(p_1,p_2)$ forms, which correspond to the  RR ground  states relevant for the index,  
are further specified by a number of requirements (see  Section 5).
Some  conditions
 enforce that the forms have the highest (lowest) weight 
 under the action  of the zero mode subalgebra
of   $\widehat{su(2)}_+ \oplus \widehat{su(2)}_-$ 
on  differential forms on $X.$ 
Other conditions  state that the $(p_1,p_2)$ forms have 
 vanishing  interior (exterior) product 
with certain vectors (1-forms) defined in terms  of
 torsion and a triplet of complex structures on the target space. 

Our strategy to reveal the geometric meaning of the large $N=4$ index 
will be to start (Section 2) with the realization of $A_{\gamma}$ algebra for the $N=4$  gauged WZW models, review (Section 3) the structure of ground states
 in a massless R-sector representation of this algebra, identify the states
contributing to the index $I_2$ (Section 4), and, finally,  find a geometric
interpretation of  these special
states (Section 5).

\section{$WZW$ models with $A_{\c}$ symmetry.}
\setcounter{equation}{0}
In this section we review the realization of the $A_{\c}$ algebra for 
the $N=4$ 
gauged WZW models \cite{SSTV,SSTVtwo,GK,VP}. 
We are interested in these models for the following reason.
In order to find  a geometric
interpretation of the large $N=4$ index we must
start from a $\sigma$-model with  $A_{\c}$ symmetry
and express $A_{\c}$ generators in terms of the
$\sigma$-model fields. At present, the $N=4$ 
gauged WZW models  are the only known
representatives of such   $\sigma$-models and we
will rely on them.

The key issue that makes the $N=4$ 
gauged WZW models special among $\sigma$-models with  $A_{\c}$ symmetry
is their  relation with the $N=4$ cosets $G/H.$
Here $G=SU(N+2)$ or $G=G'\otimes U(1)$
with $G'$  a simple group different from $SU(M).$
In the construction of
these cosets  one chooses a subgroup $H$
 in such a way that 
 $G/\left( H \times U(2)\right)$ 
is a Wolf space \cite{Wolf}. 
The simplest case $G=SU(3),\, H=\Id $ corresponds to WZW model
for the group manifold $SU(3).$

We  recall the correspondence  between 
general supersymmetric gauged WZW models and superconformal cosets in Section 2.1.
We  use the coset description of the $N=4$ gauged WZW models
to realize the generators of $A_{\c}$ algebra in Section 2.2.

\subsection{ Review of supersymmetric gauged WZW model}
Here we recall basic facts about   supersymmetric gauged WZW models
and review their description in terms of  superconformal cosets.
We follow the  discussion in \cite{FS}.

It is convenient to work in (1,1) superspace with coordinates $Z=(z,\tht)$ and
${\overline Z}=(\barz,\btht).$ Then the fields are arranged in the
superfield as:
\be
\bG=g\Bigl( 1+\tht \psi +\btht g^{-1} \bpsi g - \tht \btht \psi g^{-1}\bpsi g \Bigr)
\label{superg}
\ee
where $g(z,\barz)$ is a map from Riemann surface $\Sigma$ to the group manifold $G.$
The fermions $\psi(z,\barz)$ and $\bpsi(z,\barz)$ take values in the Lie algebra ${\bf g}$  of G.
There are also gauge superfields $\bA$ and $\barbA$ whose components take
values in the complexification of the Lie algebra ${\bf h}$ of H, a subgroup of G. 

The $\sigma$-model action has the form:
\be
S\bigl(\bG,\bA,\barbA\bigr)=\int d^2\theta  \Biggl \{
 I\bigl(\bG\bigr)+\frac{k}{4 \pi}\int_{\Sigma} d^2z Tr'\Biggl( \bA {\bG}^{-1} \barD \bG -D\bG {\bG}^{-1}\barbA -\bA \barbA+
\bA {\bG}^{-1} \barbA \bG \Biggr) \Biggr \}
\label{gWZW}
\ee
where $k\in {\bf Z}_{+}$ and $D=\p_{\tht}+\tht \p_z,\ \barD=\p_{\btht}+\btht \p_{\barz}$ 
and we denote
\be
I\bigl(\bG\bigr)=-\frac{k}{8 \pi}\Biggl \{
\int_{\Sigma} d^2z Tr'\Biggl( {\bG}^{-1} D \bG {\bG}^{-1} \barD \bG \Biggr)
+ \int_{B} dt d^2z Tr' \Biggl({\tbG}^{-1} \p_t \tbG 
\Bigl[{\tbG}^{-1} D \tbG, {\tbG}^{-1} \barD \tbG\Bigr] \Biggr) \Biggr\}
\label{WZW}
\ee
Here  $t,z,\barz$ are coordinates on  a 3-manifold $B$  such that $\p B=\Sigma$  
and $\tbG$ is an extension of the map $\bG.$ 

We use notation $Tr'=\frac{1}{x_R}Tr_R$ where   
$x_R$ is the Dynkin index of a representation R
of the Lie algebra ${\bf g}.$  The hermitean generators $T_{\hA}$ satisfy 
\be
\bigl [T_{\hA},T_{\hB}\bigr ]=if_{\hA \hB \hC}T_{\hC},
\quad Tr_R \, T_{\hA} T_{\hB}=2x_R \delta_{\hA \hB}
\label{norm}
\ee
where $f_{\hA \hB \hC}$ are anti-symmetric in all three indices. (We do not distinguish lower and upper indices.) 

 To realize $A_{\gamma}$ algebra in section 2.2 we will use
the correspondence between supersymmetric gauged  WZW models and superconformal cosets.
This relation   was established in \cite{FS} following the early proof
for the bosonic CFT \cite{Ts}.  
The idea is to fix the gauge $\barbA=0$ and  change 
variables from $\bA$  to $\bH$ by parametrizing $\bA=-D\bH {\bH}^{-1}.$
 This procedure  has to be accompanied by
 the introduction of ghost superfields $\bB,\bC,\barbB,\barbC$ taking values in the
complexification of the Lie algebra ${\bf h}$  with
the action:
\be
S_{ghost}=-\frac{k}{8 \pi}\int d^2\theta d^2z \Biggl \{
 Tr'\Bigl( {\barbB}D{\barbC}+ \bB \barD \bC\Bigr) \Biggr\}
\label{ghosts}
\ee
Also, using supersymmetric version of Polyakov-Wiegmann identity,
one may write
\be 
S'(\bG,\bH):= S(\bG,\bA=-D\bH {\bH}^{-1},0)=\int d^2\tht \Bigl( I(\bG\bH)-I(\bH)\Bigr )
\label{pol}
\ee
where $I(\bG)$ is defined in (\ref{WZW}).
The last step is to make change of variables $\bG \to \bG {\bH}^{-1}$ which has a
trivial Jacobian.
Then, the path integral takes the form:
\be
{\bf Z}=\int [d\bG][d\bH][d\bB][d\bC][d\barbB][d\barbC]e^{-\int d^2\tht \Bigl( I(\bG)-I(\bH)\Bigr )-S_{ghost}} 
\label{partition}
\ee
The key point in the proof of \cite{FS} is that the total energy momentum-tensor 
\be
{\bf T}={\bf T}_{\rm g}+ {\tilde {\bf T}}_{\rm h}+{\bf T}_{ghost}
\label{tensor}
\ee
is equal to the one of the coset CFT up to a BRST-trivial term.
Here ${\bf T}_{\rm g}$ corresponds to WZW model for the group $G,$ 
the ``tilde'' in  ${\tilde {\bf T}}_{\rm h}$ indicates
that the term with Cartan-Killing metric is taken with negative sign,   
and the ghost piece ${\bf T}_{ghost}$ has the standard form.

Let us clarify this key issue a bit more. 
The components of the term ${\bf T}_{\rm g}=\half G_{\rm g} (z)+\tht T_{\rm g}(z) $ 
 are:
\be
T_{\rm g}=\frac{1}{k}\Biggl(:\! E^{\hA}(z) E^{\hA}(z) \!:-:\! \Psi^{\hA}(z)\p \Psi^{\hA}(z) \!:\Biggr)
\label{tten}
\ee
\be
G_{\rm g}=\frac{2}{k}\Biggl( \Psi^{\hA}(z) E^{\hA}(z)-\frac{i}{3k}H_{\hA \hB \hC}
:\! \Psi^{\hA}(z) \Psi^{\hB}(z)\Psi^{\hC}(z) \!:\Biggr)
\label{gten}
\ee
where $H_{\hA \hB \hC}=\frac{1}{\sqrt{2}} f_{\hA \hB \hC} $ and the currents
$E^{\hA}(z),\, \Psi^{\hA}(z)$ are given in terms of $\sigma$-model fields
(\ref{superg}) as:
\be
E^{\hA}(z)=-\frac{k}{\sqrt{2}} \Bigl(  \p g g^{-1} \Bigr )^{\hA},\quad
\Psi^{\hA}(z)=-\frac{k}{\sqrt{2}}\Bigl( g \psi g^{-1}\Bigr)^{\hA} 
\label{mycur}
\ee
They have the following non-vanishing OPE's 
\be
E^{\hA}(z) E^{\hB}(w) \sim \frac{n}{2}\frac{ \delta^{\hA \hB}}{(z-w)^2}+ 
\frac{ iH_{\hA \hB \hC}E^{\hC}(w)}{z-w}
 \label{opex}
\ee
\be
\Psi^{\hA}(z)\Psi^{\hB}(w)\sim \frac{k}{2}\frac{\delta^{\hA \hB}}{z-w}  
\label{opepsi}
\ee
where $n=k-c_g$ and $c_g$ is dual coxter number of group G.

Now, one defines the energy-momentum tensor ${\bf T}_{coset}=\half G_{coset} (z)+
\tht T_{coset}(z) $ for the coset CFT:
\be
T_{coset}=
\frac{1}{k}:\! E^A E_A \!: +\frac{2i}{k^2} H_{ABI} E^I :\! \Psi^A \Psi^B \!:-
\frac{n}{k^2}:\! \Psi_A\p\Psi^A \!: 
\label{tcoset}
\ee
$$
- \frac{1}{k^3}H_{ABC}H^A_{~~EF}:\! \Psi^B \Psi^C\Psi^E \Psi^F \!: +
 \frac{2}{k^2} H_{ABC}H^{EBC}
:\! \Psi_E \p \Psi^A \!:
$$

\be
G_{coset}=\frac{2}{k}\Biggl( \Psi^{A}(z) E^{A}(z)-\frac{i}{3k}H_{A B C}
:\! \Psi^{A}(z) \Psi^{B}(z)\Psi^{C}(z) \!:\Biggr)
\label{gcoset}
\ee
where we split the  generators $T^{\hA}=(T^A,T^I)$ (\ref{norm}) so that 
$T^I,\ I=1,\ldots,dim(H)$  span the Lie algebra ${\bf h}$ while index ``A'' runs over the coset. 
Then, $${\bf T}_{\rm g}-{\bf T}_{coset} + {\tilde {\bf T}}_{\rm h}+{\bf T}_{ghost}$$
was shown to be a BRST-trivial operator.

\subsection{ $A_{\gamma}$ generators in terms of the gauged WZW fields.} 
In section 2.1 we reviewed coset description for general supersymmetric
gauged WZW models. Here we recall how the coset description for the $N=4$ 
gauged WZW models is used to realize the generators of the left-moving
$A_{\c}$ algebra.

The  $A_{\c}$ algebra  consists of Virasoro current $T(z),$ four
dimension $3/2$ supersymmetry generators $G^a(z),$ six  
generators $A^{\pm; i}$
 of the current algebra $\widehat{su(2)}_+ \oplus \widehat{su(2)}_- ,$
 four dimension $1/2$ fermions $Q^a(z)$ and a dimension 1 boson $U(z).$
The central charge of this algebra 
is parametrized by two natural numbers $k_+,k_-$ and has the form
$c=\frac{6k_+ k_-}{k_+ + k_-}.$ The OPE's of  $A_{\c}$ algebra  
are given in Appendix B.

The realization of  $A_{\c}$ algebra for the $N=4$ coset  starts from promoting 
the supersymmetry current (\ref{gcoset}) into 
the four supersymmetry generators. We will omit the subscript
``coset" from now on:
\be
G^{a}=\frac{2}{k }\Bigl[  \Psi^A J^{a}_{AB}E^{B}
-\frac{i}{3k}S^{a}_{ABC} :\! \Psi^A\Psi^B\Psi^C \!: \Bigr]
\label{g}
\ee
where $J^a_{AB}=(J^i_{AB},\delta_{AB}),\ a=(i,4), \ i=1,2,3.$ 
The objects  
$J^{i }_{AB}$ and $S^{a}_{ABC}$ will be
determined shortly.  

Using OPE's (\ref{opex})(\ref{opepsi}) of the currents $E^{\hA}(z),\Psi^A(z)$   we find:
$$
G^{a}(z)G^{b}(w)\sim \frac{4}{k^2}\Biggl \{ \frac{k/4}{(z-w)^3}\Bigl[nJ^{a}_{AB}  J^{b AB}
+\frac{1}{3}S^{a}_{ABC}S^{bABC} \Bigr]+
$$
$$ \frac{1}{(z-w)^2}\Bigl[\frac{n}{2}J^{a}_{AC}  J^{b~C}_{E}:\! \Psi^A \Psi^E \!: 
+\frac{ik}{2}J^{aA}_{~~~C}  J^{b}_{AE}H^{CE\hB}E_{\hB}
+\half S^{a}_{BCA}S^{bBC}_{~~~~~~E}:\! \Psi^A \Psi^E \!: \Bigr]+
$$
$$
\frac{1}{z-w}\Bigl[ \frac{k}{2} J^{a}_{CA}  J^{bC}_{~~~~B} :\! E^A E^B \!: 
+iJ^{a}_{AC}  J^{b}_{BD}H_{\hF}^{~~CD}E^{\hF} :\! \Psi^A \Psi^B \!:$$
\be
+\frac{ik}{4} J^{a}_{AC}  J^{b~A}_{~~~~E}H_{\hD}^{~~CE}\p E^{\hD} 
-i S^{(a}_{AED}J^{b) ~A}_{~~~~~B}E^B :\! \Psi^E \Psi^D \!: +
\label{gope}
\ee
$$\frac{n}{2}J^{a}_{BC}  J^{b~~C}_{A}:\! \p (\Psi^B) \Psi^A \!: 
+\half S^{a}_{BCA}S^{b~BC}_{~~~~~~E} :\! \p (\Psi^A )\Psi^E \!: 
-\frac{1}{2k} S^{a}_{ABC}S^{bA}_{~~~EF} :\! \Psi^B\Psi^C \Psi^E\Psi^F \!: \Bigr]\Biggr \}
$$

We have to compare (\ref{gope}) with the corresponding relation
in $A_{\gamma}$ algebra:
\be
G^{a}(z)G^{b}(w)\sim \frac{2c}{3} \frac{\delta^{ab} }{(z-w)^3}
-\frac{4k_- t^{+i}_{ab}A^{+i}(w)+ 4k_+ t^{-i}_{ab}A^{-i}(w)}{k(z-w)^2}
\label{compope}
\ee
$$-\frac{2k_- t^{+i}_{ab}\p A^{+i}(w)+ 2k_+ t^{-i}_{ab}\p A^{-i}(w)}{k(z-w)}
+\frac{2\delta^{ab} T(w)}{z-w} $$
where
$t^{\pm i}_{ab}=\pm 2\delta^i_{[a}\delta^4_{b]}+ \e_{iab}$ and 
$$
c=\frac{6k_{+}k_{-}}{k},\quad k=k_{+}+k_{-}
\label{c}
$$

Let us first look at the terms $\sim \frac{1}{z-w}$ which are symmetric
in indices $a,b.$ Comparing (\ref{gope}) with (\ref{compope})
gives the following constraints:
\be
J^{(a}_{AB}  J^{b)A}_{~~~~C}=\delta^{ab}\delta_{AC},\quad
J^{(a}_{AC}  J^{b)~~C}_{B}=\delta^{ab}\delta_{AB}
\label{jj}
\ee
\be
J^{(a}_{AC}J^{b)}_{BD}H^{CD}_{~~~~E}-J^{D (a}_{~~~~E} S^{b)}_{DAB}=\delta^{ab}F_{EAB}
\label{jjh}
\ee
\be
J^k_{CD}H^D_{~~BI}=J^{kD}_{~~~B}H_{CDI},\quad k=1,2,3
\label{hinv}
\ee
\be
S^{(a}_{D[ AB} S^{b)~~D}_{EF]}=\delta^{ab}P_{ABEF},\quad
S^{(a}_{ABE} S^{b)AB}_{F}=\delta^{ab}\nu_{EF}
\label{ss}
\ee
where tensors $F_{EAB}, P_{ABEF}, \nu_{EF}$ will be determined
in a moment.

The  constraints (\ref{jj}) require that $J^i_{AB}$ are almost complex structures
on the coset, i.e. they are
anti-symmetric\footnote{Recall that we do not distinguish upper and lower indices.}
\be
J^i_{AB}=-J^i_{BA}
\label{anti}
\ee
and 
satisfy the algebra of imaginary quaternions
\be
 J^{i A}_{~~~B} J^{j B}_{~~~C}=\varepsilon_{ijk}J^{k A}_{~~~C}-\delta^{ij}\delta^{A}_C
\label{quat}
\ee

Let us consider (\ref{jjh}). Taking first $a=i$ and $b=4$
we find
\be
S^{4}_{ABC}= H_{ABC},\quad S^i_{ABC}=3 J^{i~~~D}_{[A}H_{BC]D}
\label{sol}
\ee

Next, consider $a=4,b=4$ to obtain $F_{EAB}=0$ and
 $a=i,b=j$ to find
\be
3J^{~~~A (i}_{[C}J^{j)~~~B}_{D}H_{E]AB}=H_{CDE}\delta^{ij}
\label{nij}
\ee
Note that (\ref{nij}) is the condition for integrability of the complex structures $J^i.$
In \cite{SSTV} there is an explicit expression for the triplet $J^i$ satisfying 
(\ref{nij}).
From this condition follows that $S^a_{ABC}$ can be also
recast as
\be 
S^a_{ABC}= J_{A}^{a~~D_1} J_{B}^{a~~D_2}J_{C}^{a~~D_3} H_{D_1 D_2 D_3}
\label{solii}
\ee
 From (\ref{hinv})(\ref{sol}) and using also:
\be
H_{A [BC}H^A_{~~~EF]}+H_{I [BC}H^I_{~~~EF]}=0
\label{jacobiiden}
\ee
\be
H_{ABC}H^{ABD}=c_g\delta^D_C-2H_{IBC}H^{IBD}
\label{property}
\ee
we check that conditions
 (\ref{ss}) are true with
\be
 P_{ABEF}=H_{D[AB}H^D_{~~EF]},\quad \nu_{EF}=H_{ECD}H^{~~CD}_F
\label{sh}
\ee

Let us now look at the term $\sim \frac{1}{(z-w)^3}$ in 
(\ref{gope}).
From (\ref{solii}) we find
\be
S^a_{ABC}S^{bABC}=3d_X \delta^{ab}
\label{conf}
\ee
 We  used that for the $N=4$ coset $X=G/H$  
 $$c_h=c_g-2, \quad d_X=4(c_g-1),\quad dim H =c_h^2-1$$
and thus
\be
H_{ABC}H^{ABC}=\Bigl( c_gd_X-2(c_g-c_h) dim H \Bigr)=3d_X
\label{vajno}
\ee
From (\ref{conf}) we find that  the term $\sim \frac{1}{(z-w)^3}$  has the correct structure as in (\ref{compope}) with central charge:
\be
c=\frac{6k_+k_-}{k},\quad k_-=c_g-1,\quad k_+=n+1
\label{inded}
\ee
Finally, let us consider the term $\sim \frac{1}{(z-w)^2}$ in
(\ref{gope}). We use the following helpful formulae:
\be
J^{a}_{A[B}J^{bA}_{~~~C]}=-(t^{+i})^{ab}J^i_{BC},\quad
J^{[a}_{EC}J^{b]~C}_{F}=-(t^{-i})^{ab}J^i_{EF}
\label{usei}
\ee
\be
S^a_{BC[A}S^{bBC}_{~~~~E]}=-(t^{-i})^{ab}J^i_{AE}+(t^{+i})^{ab}M^{i}_{AE},\quad
M^{i}_{AE}=S^i_{BC[A}H^{BC}_{~~~E]}-J^i_{AE}
\label{useii}
\ee
to find the correct structure as in (\ref{compope}) with
\be
A^{-;i}=\frac{1}{2k}J^i_{AE}:\! \Psi^A\Psi^E \!:
\label{aminus}
\ee
\be
A^{+;i}=\frac{i}{2k_-}\Biggl[J^i_{AE}H^{AE\hD}E_{\hD} + 
\frac{i}{k} M^i_{AE}:\! \Psi^A\Psi^E \!:\Biggr]
\label{aplus}
\ee
Now let us check  the  OPE's of $\widehat{su(2)}_+ \oplus \widehat{su(2)}_- $ currents:
$$
A^{\pm;i}(z)A^{\pm;j}(w)\sim -\frac{k_{\pm}}{2}\frac{\delta^{ij}}{(z-w)^2}+
\frac{\varepsilon_{ijk}A^{\pm;k}}{z-w}
$$
\be
A^{-;i}(z)A^{+;k}(w)=reg
\label{opeaaii}
\ee
In order to prove (\ref{opeaaii}), it is convenient to bring $M^i_{AB}$   into the form
\be
M^{i}_{AB}=k_-J^i_{AB}-h^i_CH^C_{~~AB},\quad h^i_F=J^{i~AB}H_{ABF},\ i=1,2,3
\label{hif}
\ee
Note that (\ref{hif})  follows from  the general expression (\ref{useii})
by using the  property 
 of the $N=4$ cosets 
\be
J^{k~AB} H_{ABI}=0,\quad k=1,2,3
\label{specific}
\ee 
This property  is due to the specific choice of a subgroup $H$
in the construction of the $N=4$ cosets $G/H.$
 We recall that for these cosets
$G=SU(N+2)$ or $G=G'\otimes U(1)$
with $G'$  a simple group different from $SU(M).$
A subgroup $H$ is chosen in
such a way that its simple roots are orthogonal to the highest root
of $SU(N+2)$ or $G'.$ 

Then, OPE's (\ref{opeaaii}) are reproduced correctly due to the following 
properties of $h^i_F$: 
\be
h^i_Ah^j_BH^{ABC}=-2k_- \epsilon_{ijk}h^{k~C}
\label{usev}
\ee
\be
h^i_{C}h^{k~C}=4k_-^2\delta^{ik},\quad 
h^i_D h^k_F H^{DAB}H^{F}_{~~ AB}=4k_-^2(k_-+1)\delta^{ik}
\label{useiii}
\ee
Checking the rest of the OPE's of $A_{\gamma}$ algebra (see Appendix B),
determines
\be
Q^b=\frac{i}{2k_-} h^{b}_F\Psi^F
\label{qb}
\ee
\be
U=\frac{i}{2k_-}h^4_F\Bigl(E^F-\frac{i}{k}H^F_{~~ED}:\Psi^E\Psi^D:\Bigr)
\label{u}
\ee
where we denote 
\be
h^b_F=(h^i_F,h^4_F),\ i=1,2,3, \quad h^4_F=-J^{i~~C}_F h^{i}_{C} \ \forall i
\label{hbf}
\ee
and $h^{i}_F$ is defined in (\ref{hif}). 

Note that second term in (\ref{u}) is non-zero only for the case
$G=SU(N+2).$ For the other choice $G=G'\otimes U(1)$ one finds that $U$
is a generator of the $U(1)$ factor.

As an example of the realization of $A_{\c}$ algebra, we present  explicit expressions for 
$J^i_{AB},\, H_{ABC}$ and $ h^a_F,\, M^i_{AB}$
for the  case of $SU(3)$ WZW model in Appendix C.

\section{The ground states in a massless R-sector $A_{\gamma}$ module.}

\setcounter{equation}{0}
In this section we review some of the results \cite{GPTV,PTone,PT} about RR characters
of $A_{\c}$ algebra and the structure
of the ground states
in a massless R-sector $A_{\c}$ module.
This information will be used in Section 4
to identify the special states contributing to the index $I_2.$

Let ${\cal H}_{RR}$ be a representation of the RR sector algebra\footnote{We do not impose
any GSO projection. }
$A_{\gamma}^{left}\oplus A_{\gamma}^{right}.$ 
The RR sector $A_{\gamma}$ character is defined by \cite{PT}:
\be
Z(q,z_{\pm};{\overline q},{\barz}_{\pm}):=
 Tr_{{\cal H}_{RR}} z_-^{2iA_0^{-;3}}{\barz}_-^{2i{\overline A}_0^{-;3}}
z_+^{2iA_0^{+;3}} {\barz}_+^{2i{\overline A}_0^{+;3}}q^{L_0-c/24} {\overline q}^{{\overline L}_0-c/24}
\label{superch}
\ee

For a general unitary theory with $A_{\c}$ symmetry
$Z$ has a  form:
\be
Z=\sum_{u,\baru} \sum_{l_+,\barl_+=\half}^{\frac{k_+}{2} } \sum_{l_-,\barl_-=\half}^{\frac{k_-}{2} } 
N_{l_+,l_-,u;{\overline l}_+,{\overline l}_-,{\overline u} } 
  SCh_0^{A_{\gamma},\  R}(l_+,l_-,u; q, z_{\pm})  SCh_0^{A_{\gamma},\  R}({\overline l}_+,{\overline l}_-,{\overline u};
 {\overline q},{\barz}_{\pm}) 
+\ldots 
\label{partitionii}
\ee
where  the contribution of the massless R-sector representation
 $r=(l_+,l_-,u)$ is given by
\be
 SCh_0^{A_{\gamma},\  R}(l_+,l_-,u; q, z_{\pm}) =Tr^R_{r}\, z_-^{2iA_0^{-;3}}
z_+^{2iA_0^{+;3}} q^{L_0-c/24},
\label{superchirr}
\ee
and $\ldots$ stands for the massive characters which are irrelevant for us
since, as shown in \cite{GMMS},
 they do not contribute  to the index $I_2.$
 
The  $A_{\c}$ character of the module $r=(l_+,l_-,u)$ can be written in a product form: 
\be
SCh_0^{A_{\gamma},\  R} \bigl( l_{+},l_{-},u; q, z_{\pm}\bigr)=
{\cal S}^{R}\bigl(u; q,z_{\pm}\bigr) \times SCh_0^{{\tilde A}_{\gamma},\  R} 
\bigl(\l_+,\l_-; q,z_{\pm}\bigr)
\label{ch}
\ee
Here ${\cal S}^R(u;q,z_{\pm}\bigr)$ denotes  the R-sector
character of the model $S$ which is a theory of  
 four Majorana fermions  and one free boson.
The model $S$ is the simplest theory with $A_{\c}$ symmetry.

$SCh_0^{{\tilde A}_{\gamma},\  R}\bigl(\l_+,\l_-; q,z_{\pm}\bigr) $ is the  character of 
the massless R-sector representation  ${\tilde r}=(\l_+,\l_-)$ of  the non-linear
algebra $\A_{\gamma}$ \cite{GodSh}    
and $\l_{\pm}=l_{\pm}-\half.$ We give explicit expressions
for  the characters ${\cal S}^R\bigl(u;q,z_{\pm}\bigr)$ and 
$SCh_0^{{\tilde A}_{\gamma},\  R}\bigl(\l_+,\l_-; q,z_{\pm}\bigr)$
in Appendix~D. 

In what follows  we will use the expressions \cite{GodSh} of the 
generators of the non-linear algebra ${\tilde A}_{\gamma}$ in
terms of the $A_{\gamma}$ generators:  
\be
\T=T+\frac{1}{k} U^2+\frac{1}{k} \p Q^a Q^a
\label{ttilde}
\ee
\be
\G^a=G^a+\frac{2}{k}UQ^a-\frac{2}{3k^2}\epsilon_{abcd}Q^bQ^cQ^d+
\frac{2}{k}Q^b\Bigl[
t^{+i}_{ba}\A^{+;i}-t^{-i}_{ba}\A^{-;i}\Bigr]
\label{tildeg}
\ee
\be
\A^{\pm;i}=A^{\pm;i}-\frac{1}{2k}(t^{i\pm})^{ab}Q_aQ_b
\label{tildea}
\ee

The product structure (\ref{ch}) implies the decomposition of the ground states 
in the module $r=(l_+,l_-,u)$ as
\be
\vert V \rangle \otimes \vert f_s \rangle \otimes \vert u \rangle
\label{ground}
\ee
Here $\vert f_s \rangle \otimes \vert u \rangle, \, s=1,\ldots, 4 $ are R ground states of 
the model $S,$ while
$\vert V \rangle$ are the ground states in 
the R-sector module ${\tilde r}=(\l_+,\l_-)$ of 
the $\A_{\gamma}$ algebra. (By $\vert u \rangle$ we denote the ground
state of the boson and by  $\vert f_s \rangle $ the R ground states of the
four fermions in the model $S.$ ) 

As was shown in \cite{PT}, the states $\vert V \rangle$
form two irreducible representations of zero-mode algebra $\A_0^{\pm;i}.$
The first one, let us call it $\vert V_1 \rangle,$ is built by acting with  operators
$\A_0^{\pm; -}$ on the highest weight state $\vert \O \rangle$ defined by
the following equations
$$
\G_0^{+} \vert \O \rangle =0,\quad K^{\pm} \vert \O \rangle =0,
$$
\be
\A_0^{\pm;+} \vert \O \rangle  =0,\quad i\A_0^{\pm;3} \vert \O 
\rangle=\l_{\pm} \vert \O \rangle
\label{pt}
\ee
where $\l_{\pm}=l_{\pm}-\half$
and we use the notations:
\be
\G_0^{\pm}=\G_0^{1}\pm i \G_0^{2},\quad K^{\pm}= \G_0^{3}\pm i \G_0^{4} ,\quad
\A_0^{\pm;\pm}=\A_0^{\pm;1}\pm i \A_0^{\pm;2}  
\label{zerog}
\ee

The second representation $\vert V_2 \rangle$  is built by acting with  operators
$\A_0^{\pm; -}$ on the state $\G_0^{-}\vert \O \rangle$
which has the properties:
\be
\A_0^{\pm;+} \G_0^{-}\vert \O \rangle =0,\quad 
i\A_0^{\pm;3} \G_0^{-}\vert \O \rangle=(\l_{\pm}-\half )\G_0^{-}\vert \O \rangle
\label{hwii}
\ee

Note, that from unitarity and (\ref{pt}) follows:
\be
{\tilde L}_0 \vert \O \rangle =\h \vert \O \rangle ,\quad \h=\frac{(k_+ -1)(k_- - 1)}{4k}+
\frac{(\l_+ + \l_-)(\l_+ + \l_- +1)}{k}
\label{cdim}
\ee

Recalling also conformal dimensions of $\vert f_s \rangle$ and $\vert u \rangle$:
$$ h_f=\frac{1}{4},\quad h_u=\frac{u^2}{k}$$
we compute the
conformal dimension of the total R groundstate (\ref{ground}):
$$ h_{R, \ ground}=\h + h_f + h_u=\frac{c}{24}+ \frac{\mu^2}{4k}+ \frac{u^2}{k}$$ 
where  $\mu=2(l_+ + l_-)-1=2(\l_+ + \l_-)+1.$

\section{States  contributing to the index $I_2.$} 

\setcounter{equation}{0}

In this  section  we  study the structure of the  index $I_2.$
We  identify the special RR ground states whose orbits
under spectral flow generate 
 the contribution of a massless $r \otimes {\overline r}$
 module to the index. 
We will give a geometric interpretation for these  states 
in terms of differential forms on the target space in Section 5.

First, we recall some results \cite{GMMS} about the
structure of the index $I_2.$
Using (\ref{intro}) and (\ref{partitionii}) the large N=4 index $I_2$ can be written as:
\be
I_2=\sum_{r}\sum_{{\overline r}}N_{r, {\overline r}}Ind_r(q,z)Ind_{{\overline r}}({\overline q},\barz)
\label{itwo}
\ee
where $Ind_r(q,z)$ stands for the contribution of 
the left-moving massless R-sector $A_{\gamma}$ module $r=(l_+,l_-,u)$ 
\be
Ind_r(q,z):= \Biggl [ z_+ \frac{d}{dz_- } SCh_0^{A_{\gamma},\  R} \bigl( l_{+},l_{-},u; q, z_{\pm}\bigr) \Biggr]_{z_-=-z,\, z_+=z}
\label{defind}
\ee
and $D_0^3=i(A_0^{+;3}+A_0^{-;3}).$

The factorized  form (\ref{ch}) of the character 
$SCh_0^{A_{\gamma},\  R} \bigl( l_{+},l_{-},u; q, z_{\pm}\bigr)$
and explicit expressions for the two factors (see Appendix~D) allow to compute 
\be
Ind_r(q,z)=(-)^{2l_--1} q^{u^2/k} \Theta^{-}_{\mu,k}(q,z),\quad \mu=2(l_+ + l_-)-1.
\label{indcomp}
\ee 
From (\ref{itwo})(\ref{indcomp}) and the structure of the odd theta function
$\Theta^{-}_{\mu,k}(q,z)$ immediately follows 
that states in the module $r \otimes {\overline r}$ 
which contribute to the index $I_2$ satisfy:
 \be
2D_0^3 =\pm \mu +2km, \quad L_0-c/24= \frac{u^2}{k} + \frac{(D_0^3)^2}{k},\quad
m\in {\bf Z}
\label{condi}
\ee
\be
 2\barD_0^3 =\pm \barmu +2k\barm, \quad 
 \barL_0-c/24= \frac{\baru^2}{k} + \frac{(\barD_0^3)^2}{k},\quad \barm\in {\bf Z}
\label{condii}
\ee                                                                               
where $D_0^3=i\left(A^{+;3}_0+A^{-;3}_0\right),\ \barD_0^3=i\left(\barA^{+;3}_0+\barA^{-;3}_0\right)$ and $\barmu=2(\barl_+ + \barl_-)-1.$

Next, we make an important observation.
The conditions (\ref{condi})(\ref{condii})
are invariant under symmetric spectral flow  with  parameters\footnote{ Symmetric spectral flow
with parameter $\rho$ acts as
$ L_{0}^{ (\rho, \rho) }=L_0-\rho D_0^3 +\frac{k\rho^2}{4},\, D_0^{3 (\rho,\rho) }=D_0^3- 
\frac{k\rho}{2}$}
$\rho=2n,\, {\overline \rho}=2{\overline n}$:
\be                                                                     
L_{0}^{ (2n, 2n) }=L_0-2n D_0^3 +kn^2,\quad
D_0^{3 (2n,2n) }=D_0^3- kn,\quad n\in \Z 
\label{flowii}
\ee  
\be                                                                     
{\overline L}_{0}^{ (2\barn, 2\barn) }={\overline L}_0-2\barn {\overline D}_0^3 +
k\barn^2,\quad
{\overline D}_0^{3 (2\barn, 2\barn) }={\overline D}_0^3- k\barn, \quad \barn\in \Z 
\label{flowiii}
\ee 
There are 16 ground states in the  
module $r \otimes {\overline r}$ which 
satisfy   (\ref{condi})(\ref{condii}) for $m=0,\, {\overline m}=0$ and
generate 
 the contribution of this module to
the index by means of spectral flow (\ref{flowii})(\ref{flowiii}).
 
These special RR ground states are 
\be
\vert s,\bars \rangle=\vert g_s \rangle \otimes
\vert \barg_{\bars}\rangle,\quad s=1,\ldots,4,\quad \bars=1\ldots,4
\label{special}
\ee 
where $\vert g_s \rangle$ is one of 
the following  special ground states in the left-moving module                                                                  
\be
\vert g_1\rangle =\vert \O \rangle \otimes  \vert f_1 \rangle\otimes \vert u\rangle,\quad
\vert g_2 \rangle 
=\vert \O \rangle \otimes  \vert f_2 \rangle \otimes \vert u \rangle
\label{gsi}
\ee 
\be
\vert g_3\rangle =
\Bigl(i\A^{+;-}\Bigr)^{2\l_+} \Bigl(i\A^{-;-}\Bigr)^{2\l_-}\vert \O \rangle \otimes 
 \vert f_3 \rangle \otimes \vert u \rangle,\quad \vert g_4 \rangle=
\Bigl(i\A^{+;-}\Bigr)^{2\l_+} \Bigl(i\A^{-;-}\Bigr)^{2\l_-}\vert \O \rangle
\otimes  \vert f_4 \rangle \otimes \vert u\rangle
\label{gsiii}
\ee 
In (\ref{gsi})(\ref{gsiii}) the state $\vert \O \rangle$ is defined in (\ref{pt}) and 
 R ground states $\vert f_s \rangle$ of the free fermions
are specified as
\be
\left(Q_0^1+iQ_0^2\right)\vert f_s\rangle=0,\quad s=1,2,\quad
\left(Q_0^1-iQ_0^2\right)\vert f_s\rangle=0,\quad s=3,4
\label{newferm}
\ee
$$ \left(Q_0^3+iQ_0^4\right)\vert f_s\rangle=0,\quad s=1,4,\quad
\left(Q_0^3-iQ_0^4\right)\vert f_s\rangle=0,\quad s=2,3 $$
The four right-moving states $\vert \barg_{\bars} \rangle$
which enter the definition of $\vert s,\bars \rangle $
have analogous form.

The states $\vert s,\bars \rangle$ have conformal dimensions
\be
  h_{R, \, ground}=\frac{c}{24}+ \frac{\mu^2}{4k}+ \frac{u^2}{k},\quad
 \barh_{R, \, ground}=\frac{c}{24}+ \frac{\barmu^2}{4k}+ \frac{\baru^2}{k}
\label{confdim}
\ee 
and the following $i A_0^{\pm;3}$ eigenvalues 
$$ m^{(1)}_+=l_+,\quad m^{(1)}_-=l_- -\half,\quad
m^{(2)}_+=l_+-\half,\quad m^{(2)}_-=l_-,$$
$$ m^{(3)}_+=-l_+,\quad m^{(3)}_-=-(l_- -\half),\quad
m^{(4)}_+=-(l_+-\half),\quad m^{(4)}_-=-l_- $$
with analogous expressions for $i {\overline A}_0^{\pm;3}.$ 

\section{Geometric interpretation of the index $I_2.$}  

\setcounter{equation}{0}

In the previous section  we considered 
the contribution to the index $I_2$  from massless RR-sector
$A_{\c}$ module $r \otimes {\overline r}.$ 
We found that this contribution 
comes from spectral flow
orbits of  the  special RR ground states $\vert s,\bars \rangle=\vert g_s \rangle \otimes
\vert \barg_{\bars}\rangle.$

Here we first clarify the geometric meaning  of the
left-moving  states $\vert g_s \rangle.$ Then,
we combine left and right moving sectors and 
give a geometric interpretation of the index $I_2.$

In Section 5.1 we characterize  the  states $\vert g_s\rangle$
 as Dirac spinors on the target space $X$ specified by  a system
of differential and algebraic equations (5.19)-(5.32).

In Section 5.2 we  
describe the special  RR states $\vert s,\bars \rangle$
in terms of 
differential forms on $X$ defined  in (5.36)-(5.55). 
 
In Section 5.3 we show  that  coefficients
of  leading terms in the index $I_2$ are sums
$$\sum_{p_1,p_2} (-)^{p_1+p_2}n(p_1,p_2)$$
where  $n(p_1,p_2)$ is the number of $(p_1,p_2)$ forms that  
solve equations (5.61)-(5.70). 
 
This is similar in spirit to the  geometric description \cite{KYY} of the
leading contribution  to the elliptic genus but the forms counted in the
index $I_2$
are more special. In particular, they have either the highest or the lowest
weight  under the action of $su(2)_+\oplus su(2)_-$ algebra on
differential forms on the target space.

\subsection{Geometric meaning of the special R ground states $\vert g_s\rangle.$}
Here we give a  geometric description  of the
 states $\vert g_s \rangle,$ which
are left-moving parts of the special RR ground states
found in Section 4. 

To clarify the geometric meaning of the states $\vert g_s\rangle,$  we proceed
in the following way. First, we 
 find the constraints on  $\vert g_s\rangle$ in terms of 
the  zero modes  of 
$A_{\gamma}$ generators. 
Then, by representing the zero modes   in terms of vector fields on the target
space $X$ and Dirac matrices, we 
recast the constraints on $\vert g_s\rangle$ as a system of differential and algebraic
equations on a Dirac spinor on $X.$

From the definition (\ref{gsi}) of the states $\vert g_s\rangle$ for $s=1,2$
in terms of  $\vert \O\rangle,\, \vert f_s\rangle$ 
we find 
\be
 \G_0^{+}\vert g_s\rangle=0,\quad \G_0^{3}\vert g_s\rangle=0,
\quad  \G_0^{4}\vert g_s\rangle=0, \quad s=1,2
\label{kil}
\ee
and
\be
\left(Q_0^1+iQ_0^2\right)\vert g_s\rangle=0, \quad s=1,2,\quad
\left(Q_0^3+iQ_0^4\right)\vert g_1\rangle=0,\quad \left(Q_0^3-iQ_0^4\right)\vert g_2\rangle=0.
\label{fermkil}
\ee
 As a consequence of (\ref{kil})(\ref{fermkil}), $\vert g_s \rangle$ for $s=1,2$ 
also satisfy 
\be
A_{0}^{\pm;+}\vert g_s\rangle=0, \quad G_{0}^{+}\vert g_s\rangle=0,\quad s=1,2
\label{kilnew}
\ee
where we denote
$$A_{0}^{\pm;+}=A_0^{\pm;1}+iA_0^{\pm;2}, \quad G_{0}^{+}:=G_0^1+iG_0^2. $$

Analogously, states $\vert g_s\rangle$ for $s=3,4$   (\ref{gsiii}) 
are specified by 
\be
 \G_0^{-}\vert g_s\rangle=0,\quad \G_0^{3}\vert g_s\rangle=0,
\quad  \G_0^{4}\vert g_s\rangle=0  
\label{kilii}
\ee
and
\be
\left( Q_0^1-iQ_0^2\right)\vert g_s\rangle=0,\quad s=3,4,\quad
\left(Q_0^3+iQ_0^4\right)\vert g_4\rangle=0,\quad \left(Q_0^3-iQ_0^4\right)\vert g_3\rangle=0.
\label{fermkilii}
\ee
 As a consequence of (\ref{kilii})(\ref{fermkilii}), $\vert g_s\rangle$ for $s=3,4$ also satisfy
\be
A_{0}^{\pm;-}\vert g_s\rangle=0, \quad G_{0}^{-}\vert g_s\rangle=0,\quad s=3,4
\label{kilnewii}
\ee
where 
$$A_{0}^{\pm;-}:=A_0^{\pm;1}-iA_0^{\pm;2}, \quad G_{0}^{-}:=G_0^1-iG_0^2. $$

In order to understand the geometric meaning of the constraints (\ref{kil})-(\ref{kilnewii}) 
on the states $\vert g_s\rangle$
we will represent zero modes of the operators of the $A_{\c}$ algebra
in geometric terms.
We need zero modes of the currents $E_{\hA}(z)$ and $\Psi_A(z),$ which were used in the realization
of $A_{\c}$ generators in section 2.2.
 From the OPE (\ref{opex})  we find commutation relation for zero modes
\be
\bigl[ E_{0~\hA}, E_{0~\hB} \bigr]=iH_{\hA \hB \hC} E_{0~\hC}
\label{com}
\ee
so that  $E_{0~\hA}$ can be represented in terms of
 vector field  on the group manifold $G.$
\be
E_{0~\hA}=-\frac{i\sqrt{k}}{2}E^{\hM}_{\hA} \p_{\hM}
\label{ezero}
\ee
Here  $X^{\hM}$ are coordinates on $G$ and $ E^{\hM}_{\hA}$ is the inverse of the
veilbein (see Appendix A).
It is possible to  choose the  coordinates $X^{\hM}=(x^M,\phi^m)$ in such a way that
$x^M,\ M=1,\ldots 4k_-$ parametrize the right coset space $X$ 
and $\phi^m,\ m=1,\ldots, dim H$  are coordinates on $H.$ 
Then,  $E_{A}^m=0$ and  
 $E_{0~A},\ A=1,\ldots 4k_-$ are  vector fields on  $X.$

Note that $E_{0~A}$ do not form a closed Lie algebra 
but, using the complex structures $J^i_{AB},$ there are three different ways  to split them into the two subsets which do so. 
We choose $J^3_{AB}$ and divide the generators $T_A, A=1,\ldots 4k_-$ into two groups
$T_{\a}$  and $T_{2k_- + \a}$ such that $J^3_{\a \ 2k_-+\a}=1 \quad \forall \a=1\ldots 2k_-$
Then, complex vector fields 
\be
V_{\mbox{{\boldmath $\scriptstyle \a$}}}=E_{0~2k_- +\a}-iE_{0~\a}, 
\quad V_{\albar}=E_{0~2k_- +\a}+iE_{0~\a}
\label{eigen}
\ee
form  Lie algebras ${\cal N}^{\pm}$:
\be
\bigl[ V_{\al}, V_{\bel}\bigr ]={\cal F}_{ \al \bel }^{~~~\cel} V_{\cel},\quad 
\bigl[ V_{\albar}, V_{ \belbar}\bigr ]={\cal F}_{ \albar \belbar }^{~~~\celbar} V_{ \celbar}
\label{liealgebra}
\ee
with structure constants:
\be
{\cal F}_{\albar \belbar}^{~~~\celbar}=-H_{\a \ 2k_- +\b \ 2k_-+\c}+ H_{\b \ 2k_- +\a \ 2k_-+\c}
+i H_{\a \ 2k_- +\b\  \c}-iH_{\b \ 2k_- +\a \ \c}
\label{stct}
\ee
$$ {\cal F}_{ \al \bel}^{~~~\cel}=-\Bigl( {\cal F}_{ \albar \belbar}^{~~~ \celbar} \Bigr )^*$$

Next, we  realize zero modes of fermions $\Psi^A$ in terms of hermitean Dirac matrices on  $X$
\be
\Psi_0^A= \frac{\sqrt{k} }{2} \Gamma^A, \ A=1,\ldots,4k_-
\label{dirac}
\ee
and define their  complex linear combinations:
\be
{\bf \gamma}^{\al}=\half \Bigl( \Gamma^{2k_- +\a}+ i \Gamma^{\a}\Bigr),\quad
{\bf \gamma}^{\albar}=\half \Bigl( \Gamma^{2k_- +\a}- i \Gamma^{\a}\Bigr) \quad
\{ \gamma^{\al},\gamma^{\belbar}\}=\delta^{\al \ \belbar}
\label{newgamma}
\ee
\be
\rho^{ \al}=iJ^{-}_{\albar \belbar}\gamma^{\belbar},\quad  
\rho^{\albar}=-iJ^+_{\al \bel}\gamma^{\bel}
\label{rhomatr}
\ee
where $J^{+}_{\al \bel}= J^1_{\a \b} + iJ^2_{\a \b}$ and
 $J^{-}_{\albar \belbar}= J^1_{\a \b} - iJ^2_{\a \b}$ 

The above representation (\ref{ezero})-(\ref{rhomatr}) of $E_0^A,\Psi_0^A$
 allows us to express zero modes of the supersymmetry currents $G^a(z)$ (\ref{g})
in terms of vector fields $V_{\al},V_{\albar}$ and
Dirac matrices as:
\be
G_0^{3}+i G_0^{4}=\frac{2i}{\sqrt{k}}
\Biggl[ { \gamma}^{\albar}V_{ \albar} +\half {\cal F}_{ \albar \belbar \del} { \gamma}^{\del}
{ \gamma}^{\albar \belbar } \Biggr]
\label{plusg}
\ee
\be
G_0^{3}-i G_0^{4}=-\frac{2i}{\sqrt{k}}
\Biggl[ {\bf \gamma}^{\al}V_{ \al} +\half {\cal F}_{\al \bel \delbar}  { \gamma}^{\al \bel }{ \gamma}^{\delbar} \Biggr]
\label{plusgii}
\ee
\be
G_0^{1}+i G_0^{2}=\frac{2i}{\sqrt{k}}
\Biggl[ {\rho}^{\albar }V_{\albar} +\half {\cal F}_{\albar \belbar \del} { \rho}^{\del}
{ \rho}^{\albar \belbar } \Biggr]
\label{plusgv}
\ee
\be
G_0^{1}-i G_0^{2}=-\frac{2i}{\sqrt{k}}
\Biggl[ { \rho}^{\al}V_{\al} +\half {\cal F}_{\al \bel \delbar }  { \rho}^{\al \bel }
{ \rho}^{\delbar} \Biggr]
\label{plusgiv}
\ee
Here and below the indices are lowered with $\delta_{\albar \bel}$ or
$\delta_{\al \belbar}$ and raised with $\delta^{\al \belbar}$ or $\delta^{\albar \bel}.$

Let us now find the geometric description of the
special states $\vert g_s\rangle$ (\ref{gsi})(\ref{gsiii})
relevant for the index. 
For this purpose, we express the constraints (\ref{kil})-(\ref{kilnewii})
on $\vert g_s\rangle$ in terms of vector fields $V_{\al},V_{\albar}$ (\ref{eigen}) on $X$ and
Dirac matrices (\ref{newgamma}),(\ref{rhomatr}). 
In this way the  states $\vert g_s\rangle,\ s=1,\ldots 4$ are considered as Dirac spinors on $X$ 
defined by the following system of equations.
 
I.~~~~ For all $s=1,\ldots 4$ the spinors $\vert g_s\rangle$ satisfy Dirac-like
equations constructed out of Dirac matrices $\gamma^{\al},\gamma^{\albar}$ :
\be
\Biggl[ {\bf \gamma}^{\albar}\Bigl( V_{ \albar}-\nu_{\albar}^{(s)}\Bigr) +
\half {\cal F}_{ \albar \belbar \del} { \gamma}^{\del}
{ \gamma}^{\albar \belbar } \Biggr]\vert g_{s}\rangle=0
\label{geomii}
\ee
\be
\Biggl[ {\bf \gamma}^{\al}\Bigl(V_{ \al}-\nu_{\al}^{(s)} \Bigr)
+\half {\cal F}_{\al \bel \delbar }  { \gamma}^{\al \bel }{ \gamma}^{\delbar} \Biggr]\vert g_{s}
\rangle=0
\label{plusgiii}
\ee
where
\be
\nu_{\albar}^{(s)}=-\frac{i}{2k_-}
\bart_{\albar}\bigl(\epsilon^{(s)}-i u\bigr),\quad \nu_{\al}^{(s)}=\frac{i}{2k_-}
t_{\al}\bigl(\epsilon^{(s)}+i u\bigr)
\qquad
\epsilon^{(s)}=m_+^{(s)} + m_{-}^{(s)} 
\label{nual}
\ee
and
\be
t_{\al}=h^{3}_{\a}+ i h^{3}_{ 2k_- +\a},\quad
\bart_{\albar}=\left( t_{\al}\right)^*,\quad \a=1,\ldots, 2k_-, 
\quad h^3_A=J^{3 ~ BC} H_{ABC}
\label{collin}
\ee

II.~~~~ For different $s=1,\ldots, 4$ there are  different algebraic
constraints on the spinors $\vert g_s\rangle$ :

\be
x_{\al}\gamma^{\al} \vert g_s\rangle=0,\quad s=1,2 \quad
\barx_{\albar}\gamma^{\albar} \vert g_s\rangle=0,\quad s=3,4
\label{qq}
\ee 
\be
t_{\al}\gamma^{\al} \vert g_s\rangle=0,\quad s=2,3 \quad
\bart_{\albar}\gamma^{\albar} \vert g_s\rangle=0,\quad s=1,4
\label{qqferm}
\ee 
where 
\be
x_{\al}=h^1_{\a}+ i h^2_{\a},\quad 
\barx_{\albar}=\left( x_{\al}\right)^*
\label{collinii}
\ee

III.~~~~ For $s=1,2 \, \left( s=3,4 \right)$ the spinors  $\vert g_s\rangle$ have the highest (lowest) weight
under the action of $su(2)_+\oplus su(2)_-$ algebra: 

\be
 J^+_{\al \bel}\gamma^{\al \bel} \vert g_{s}\rangle=0,\quad 
\bary^{ \delbar}V_{\delbar}\vert g_{s}\rangle=0, \quad s=1,2 
\label{anullplus}
\ee
\be
J^{-}_{ \albar \belbar}\gamma^{\albar \belbar}\vert g_{s}\rangle=0,\quad
 y^{\del}V_{ \del}\vert g_{s}\rangle=0,\quad s=3,4
\label{anullplusii}
\ee
where $y^{\del}=\barx_{\delbar},\quad \bary^{ \delbar}=x_{\del}.$

The corresponding weights are 
$iA_0^{\pm;3}\vert g_s\rangle=m^{(s)}_{\pm}\vert g_s\rangle.$
These eigenvalue equations lead to an algebraic 
constraint  on the spinors
\be 
\half \Bigl( -k_-+ \sum_{\al=1}^{2k_-}\gamma^{\al}\gamma^{\albar}\Bigr) \vert g_s
\rangle=m_{-}^{(s)} \vert g_s\rangle
\label{a3minus}
\ee
and specify the action of the vector field $f^{\al}V_{\al}-\barf^{\albar}V_{\albar}$ 

\be
\frac{i}{4k_-}\Biggl [\barf^{\delbar}V_{ \delbar}-f^{\del}V_{\del}
+\Bigl( \barf^{\delbar}{\cal F}_{\delbar \, \belbar \al }
+f^{\del}{\cal F}_{\del \al \belbar } \Bigr)
\gamma^{ \al}\gamma^{\belbar}\Biggr]\vert g_{s}\rangle
=\bigl(\frac{k_-}{2}+m_{+}^{(s)}+m_{-}^{(s)}\bigr) \vert g_s\rangle
\label{a3plus}
\ee
where $f^{\al}=\bart_{\albar},\quad \barf^{\albar}=t_{\al}.$

IV.~~~~ The spinors $\vert g_s\rangle$ satisfy one more  Dirac-like 
equation constructed out of Dirac matrices $\rho^{\al},\rho^{\albar}$
\be
\Biggl[ {\bf \rho}^{\albar} V_{ \albar} +\half {\cal F}_{\albar \belbar \del} { \rho}^{\del}
{ \rho}^{\albar \belbar } \Biggr]\vert g_{s}\rangle=0  \quad s=1,2
\label{geom}
\ee

\be
\Biggl[ {\bf \rho}^{\al}V_{\al}
+\half {\cal F}_{\al \bel \delbar }  { \rho}^{\al \bel }{ \rho}^{\delbar} \Biggr]\vert g_{s}
\rangle=0
\quad s=3,4
\label{geomiii}
\ee

V.~~~~ For all $s=1,\ldots 4$ the spinors  $\vert g_s\rangle$ have eigenvalue $u$ under the action
of $u(1)$ generator $iU.$ 
\be
-\frac{1}{4k_-}\Biggl [\barf^{\delbar}V_{ \delbar}+f^{\del}V_{\del}
+\Bigl( \barf^{\delbar}{\cal F}_{\delbar \, \belbar \al }
-f^{\del}{\cal F}_{\del \al \belbar } \Bigr)
\gamma^{ \al}\gamma^{\belbar}\Biggr]\vert g_{s}\rangle
=u\vert g_s\rangle
\label{uuplus}
\ee

\subsection{Geometric description  of the special RR ground states $\vert s,\bars \rangle$.}
In Section 4 we identified the RR ground states $\vert s, \bars\rangle$  which
generate the contribution  to the index $I_2$ 
of the massless $A_{\c}$ module $r\otimes {\overline r}.$
Here we find a
 geometric description of  these  special states
in terms of $(p_1,p_2)$ forms on the target space X.

In order to characterize  the  special states $\vert s, \bars\rangle$ 
we use  the left-moving $A_{\c}$ operators written in geometric terms in  Section 5.1
and supply analogous  expressions for the right moving ones.
Before we formulate our geometric description of the states $\vert s, \bars\rangle$,
let us clarify  the new ingredients of the construction.

The geometric realization of the four supersymmetry generators $\barG_0^{~a}$
 in the right-moving sector
involves  vector fields $W_{\al},W_{\albar}$ defined as follows. 
\be
W_{\mbox{{\boldmath $\scriptstyle \a$}}}=\barE_{0~2k_- +\a}-i\barE_{0~\a}, 
\quad W_{\albar}=\barE_{0~2k_- +\a}+i\barE_{0~\a},\quad \a=1,\ldots, 2k_-
\label{eigenclosedii}
\ee
Here $\barE_{0~A},\quad A=1,\ldots, 4k_-$ are zero modes of 
 the currents
$\barE_{\hA}(\barz)=\frac{k}{\sqrt{2}} \Bigl( g^{-1}{\overline {\p}}g \Bigl)_{\hA}.$
They are represented as vector fields on the group manifold $G$ as:
\be
\barE_{0\ \hA}=-\frac{i\sqrt{k}}{2}\barE^{\hM}_{\hA} \p_{\hM},\quad  \hA=1,\ldots, dim G
\label{ezeroclosed}
\ee
In (\ref{ezeroclosed}) $\barE^{\hM}_{\hA}$ is the inverse of the matrix
 $\barE_{\hM}^{\hA}=-i\sqrt{\frac{k}{2}}\Bigl( g^{-1}\p_{\hM} g \Bigl)^{\hA}$ and 
$g$ is a parametrization of the group manifold $G.$
We also recall that in Section 5.1 we have chosen a parametrization 
in such a way that  $x^M,\, M=1,\ldots 4k_-$
are coordinates on the right coset $X$ and $\phi^m,\, 1,\ldots, dim H$ are coordinates on the
subgroup $H.$ In this way $\barE^{m}_{A}=0$ and $\barE_{0\ A}$ become  vector fields on $X.$

The other new ingredient, that appears when
we combine left and right sectors,
are Dirac matrices $\bargamma^{\al},\bargamma^{\albar}$ 
which represent zero modes of the right-moving fermions.

Now we give a geometric interpretation of the states $\vert s,\bars>$
relevant for the index $I_2$
in terms of differential forms on the target space $X$: 
\be
\vert s,\bars \rangle=\frac{1}{p_1!}\frac{1}{p_2!}
C_{\al_1 \ldots \al_{p_1}\belbar_1 \ldots \belbar_{p_2} }^{(s,\bars)}
 \gamma^{\al_1 \ldots \al_{p_1}}\bargamma^{\belbar_1 \ldots \belbar_{p_2}}\vert hw>,
\quad \gamma^{\albar}\vert hw\rangle=0,\quad \bargamma^{\al}\vert hw\rangle=0
\label{closed}
\ee
Here $C_{\al_1 \ldots \al_{p_1}\belbar_1 \ldots \belbar_{p_2} }^{(s,\bars)}$ are components
of  a 
 $(p_1,p_2)$ form $C_{(p_1, p_2) }^{(s,\bars)}$ 
on $X$ (with respect to  a basis of veilbein 1-forms). 

For all $s,\bars=1,\ldots,4$ the forms $C_{(p_1, p_2) }^{(s,\bars)}$ satisfy  
\be
{\cal D} C^{(s,\bars)}_{(p_1,p_2)}= \nu^{(s)} \wedge C^{(s,\bars)}_{(p_1,p_2)},
\quad \quad {\cal D}^{\dg} C^{(s,\bars)}_{(p_1,p_2)}=i_{\mw^{(s)}} C^{(s,\bars)}_{(p_1,p_2)}
\label{closediii}
\ee
\be
{\barcalD} C^{(s,\bars)}_{(p_1,p_2)}= \barnu^{(\bars)} \wedge C^{(s,\bars)}_{(p_1,p_2)},\quad 
\quad {\barcalD}^{\dg} C^{(s,\bars)}_{(p_1,p_2)}=i_{\barmw^{(\bars)}}C^{(s,\bars)}_{(p_1,p_2)} 
\label{closedv}
\ee
where symbol $i_{\displaystyle { w}}$ denotes the interior product with vector ${ w}$
and
  (anti-)holomorphic differential $\left( {\barcalD} \right) \, {\cal D}$
 acts on a $(p_1,p_2)$ form as follows:
$$\Bigl( {\cal D} C_{(p_1,p_2)}\Bigr)_{\al_1 \ldots \al_{p_1+1} \belbar_1 \ldots \belbar_{p_2} }:=
V_{[\al_1} \Bigl( C_{\al_2 \ldots \al_{p_1+1}]\, \belbar_1 \ldots \belbar_{p_2} }\Bigr)+
\frac{p_1}{2}{\cal F}^{ ~~~~~\del}_{[\al_1 \al_2}C_{|\del|\al_3 \ldots \al_{p_1+1}]\, \belbar_1 \ldots \belbar_{p_2} }$$
\be
\Bigl( {\cal D}^{\dg} C_{(p_1,p_2)}\Bigr)_{\al_1 \ldots \al_{p_1-1}\belbar_1 \ldots \belbar_{p_2} }:=
V^{\bel} \Bigl( C_{\bel \al_1 \ldots \al_{p_1-1}\, \belbar_1 \ldots \belbar_{p_2} }\Bigr)-
\frac{p_1-1}{2}
{\cal F}^{ \del \cel}_{~~~~[\al_1} C_{\al_2 \ldots \al_{p_1-1}] \del \cel \, \belbar_1 \ldots \belbar_{p_2} }
\label{differ}
\ee
$$
\Bigl( {\barcalD} C_{(p_1,p_2)}\Bigr)_{\belbar_1 \ldots \belbar_{p_2+1}\al_1 \ldots\al_{p_1}}:=
W_{[\belbar_1} \Bigl( C_{\belbar_2 \ldots \belbar_{p_2+1}]\, \al_1 \ldots\al_{p_1}}\Bigr)+
\frac{p_2}{2}{\cal F}^{ ~~~~~\delbar}_{[\belbar_1 \belbar_2}C_{|\delbar|\belbar_3 \ldots
 \belbar_{p_2+1}]\, \al_1 \ldots\al_{p_1}}
$$
$$
\Bigl( {\barcalD}^{\dg} C_{(p_1,p_2)}\Bigr)_{\belbar_1 \ldots \belbar_{p_2-1}\al_1 \ldots\al_{p_1}}:=
W^{\delbar} \Bigl( C_{\delbar \, \belbar_1 \ldots \belbar_{p_2-1}\, \al_1 \ldots\al_{p_1}}\Bigr)-\frac{p_2-1}{2}{\cal F}^{ \delbar \celbar}_{~~~~[\belbar_1} 
C_{\belbar_2 \ldots \belbar_{p_2-1}] \delbar \celbar \, \al_1 \ldots\al_{p_1}}
$$
Here  ${\cal F}_{\al \bel }^{~~~\del}$ (\ref{stct}) is torsion and vector fields
 $V_{\al},V_{\albar},W_{\al},W_{\albar}$ are defined in (\ref{eigen}),(\ref{eigenclosedii}).

The 1-forms $\nu^{(s)},\, \barnu^{(\bars)}$
and vectors $\mw^{(s)},\, \barmw^{(\bars)}$ have the following
 flat components:
\be
\nu_{\al}^{(s)}=\frac{i}{2k_-}
t_{\al}\bigl(\epsilon^{(s)}+i u\bigr),
\quad 
\barnu^{(\bars)}_{\albar}=-\frac{i}{2k_-}f^{\al}
\bigl({\overline \epsilon}^{(\bars)}-i \baru\bigr)
\label{closednuii}
\ee
\be
\mw^{(s)~\al}=-\frac{i}{2k_-}f^{\al}
\bigl(\epsilon^{(s)}-iu\bigr), \quad
\barmw^{(\bars)~\albar}=\frac{i}{2k_-}
t_{\al}\bigl({\overline \epsilon}^{(\bars)}+i \baru\bigr)
\label{closednu}
\ee
where $\epsilon^{(s)}=m_+^{(s)} + m_{-}^{(s)}, \quad {\overline \epsilon}^{(\bars)}=\barm_+^{(\bars)} + \barm_{-}^{(\bars)}$ and $t_{\al}=h^3_{\a}+ih^3_{2k_-+\a},
\quad \mf^{\al}=h^3_{\a}-ih^3_{2k_-+\a}.$
We recall also the definition of $h^i_A=(h^i_{\a}, h^i_{2k_- +\a})$
in terms of torsion and  complex structures
$h^i_F=J^{i ~ AB} H_{ABF}.$

The    degrees of the forms $C_{(p_1, p_2)}^{(s, \bars)}$ describing
RR ground states $\vert s,\bars>$
are given by:
\be
p_1^{(s)}=k_-+2m_-^{(s)},\quad p_2^{(s)}=k_- - 2{\barm}_-^{(s)}
\label{closeddegrees}
\ee

The structure of the  form  $C_{(p_1, p_2)}^{(s, \bars)}$
is further constrained as
\be
C^{(s,\bars)}_{(p_1, p_2)}= *_{(J^-,\, J^+)} B^{(s)}_{(2k_- -p_1, 2k_- -p_2 )}
\label{cversusbclosed}
\ee
where $B^{(s,\bars)}_{(2k_- -p_1, 2k_- -p_2 )}$ is a
$(2k_--p_1, 2k_--p_2)$ form which is in the kernel of ${\cal D}^{\dg}$ for $s=1,2$ 
$\Bigl( {\cal D}$ for $s=3,4 \Bigr )$ as well as in the kernel of ${\barcalD}$ for 
$\bars=1,2$ 
$\Bigl( {\barcalD}^{\dg}$ for $\bars=3,4\Bigr).$ 

The symbol 
 $*_{(J^-,\, J^+)}$ stands for Hodge-like duality operation
which maps $(n_1,n_2)$ forms into $(2k_- -n_1,2k_- - n_2)$ forms and is defined in components as

\be
\left( *_{(J^-,\, J^+)} B_{(n_1,n_2)} \right)_{\al_1 \ldots \al_{2k_--n_1}\,
\belbar_1 \ldots \belbar_{2k_--n_2}}=\kappa_{n_1}\kappa_{n_2}
B_{\del_1 \ldots \del_{n_1}\delbar_1 \ldots \delbar_{n_2}}\times
\label{closedvii}
\ee
$$J^{-\ \del_1 \bel_1}\ldots J^{- \ \del_{n_1} \bel_{n_1}}
J^{+\ \delbar_1 \belbar_1}\ldots J^{+ \ \delbar_{n_2} \belbar_{n_2}}
\epsilon_{\bel_1\ldots \bel_{n_1}\al_1 \ldots \al_{2k_- -n_1} }
\epsilon_{\belbar_1\ldots \belbar_{n_2}\albar_1
\ldots \albar_{2k_- -n_2}}
$$
where
$\kappa_{n_1}=\frac{(-)^{n_1+1}(i)^{n_1}}{n_1!},\quad 
\kappa_{n_2}=\frac{(-)^{n_2+1}(-i)^{n_2}}{n_2!}$ and
$J^{-~\al \bel}=J^1_{\a \b}-i J^2_{\a \b}, \quad
 J^{+~\albar \belbar}=J^1_{\a \b}+i J^2_{\a \b}.$

The forms $C_{(p_1,p_2)}^{(s,\bars)}$ are further specified by
a number of equations. 
 Below we list these constraints
for different values of $s,\bars,$ i.e. for the forms describing different
states $\vert s,\bars >$ (\ref{closed}) :\\
$s=1,2$ 
\be
x \wedge C_{(p_1,p_2)}^{(s,\bars)}=0
\label{newclose}
\ee
\be
J^+_{[ \al \bel} C^{(s,\bars)}_{\al_1 \ldots \al_{p_1}]\, \belbar_1 \ldots \belbar_{p_2} }=0,
\quad
\bary^{\delbar} V_{\delbar}\bigl( C^{(s,\bars)}_{\al_1 \ldots \al_{p_1}\, \belbar_1 \ldots \belbar_{p_2} }\bigr)=0
\label{closedxx}
\ee
$s=3,4$
\be
i_{\my}C_{(p_1,p_2)}^{(s,\bars)}=0
\label{newcloseii}
\ee
\be
J^{- \  \al \bel} C^{(s,\bars)}_{\al \bel \al_3 \ldots \al_{p_1} \, \belbar_1 \ldots \belbar_{p_2}   }=0, \quad
y^{\del}  V_{\del}\bigl( C^{(s,\bars)}_{\al_1 \ldots \al_{p_1}\, \belbar_1 \ldots \belbar_{p_2}  }\bigr)=0
\label{closedxix}
\ee
$\bars=1,2$
\be
i_{\barmy}C_{(p_1,p_2)}^{(s,\bars)}=0
\label{newcloseiii}
\ee
\be
J^{+ \  \albar \belbar} C^{(s,\bars)}_{\albar \belbar \, \belbar_3 \ldots \belbar_{p_2} \, \al_1 \ldots \al_{p_1}   }=0, \quad
\bary^{\delbar}W_{\delbar}\bigl( C^{(s,\bars)}_{\belbar_1 \ldots \belbar_{p_2}\, \al_1 \ldots \al_{p_1}  }\bigr)=0
\label{closedxxxii}
\ee
$\bars=3,4$
\be
\barx \wedge C_{(p_1,p_2)}^{(s,\bars)}=0
\label{newcloseiiii}
\ee
\be
J^-_{[ \albar \belbar} C^{(s,\bars)}_{\belbar_1 \ldots \belbar_{p_2}]\, \al_1 
\ldots \al_{p_1} }=0,
\quad 
y^{\del} W_{\del}\bigl( C^{(s,\bars)}_{\belbar_1 \ldots \belbar_{p_2}\, \al_1 \ldots \al_{p_1} }\bigr)=0
\label{closedxxxvi}
\ee
\vspace{0.5mm}

\be
i_{\mf}C_{(p_1,p_2)}^{(s,\bars)}=0 \quad \mbox{for}\quad s=1,4,
 \quad t\wedge C_{(p_1,p_2)}^{(s,\bars)}=0 \quad \mbox{for} \quad
 s=2,3,
\label{finita}
\ee
\be
\bart \wedge C_{(p_1,p_2)}^{(s,\bars)}=0 \quad \mbox{for} \quad \bars=1,4,\quad 
i_{\displaystyle \! {\bar f}}C_{(p_1,p_2)}^{(s,\bars)}=0 \quad \mbox{for} \quad \bars=2,3
\label{finitaiii}
\ee
where the flat components of the 1-forms $x,\, \barx,\, t, \, \bart $ and vectors
$y,\, \bary,\, f,\, \barf$ are given by
\be
x_{\al}=h^1_{\a}+ih^2_{\a}, \quad t_{\al}=h^3_{\a}+ih^3_{2k_-+\a},\quad
\my^{\al}=h^1_{\a}-ih^2_{\a}, \quad \mf^{\al}=h^3_{\a}-ih^3_{2k_-+\a}
\label{flatclosed}
\ee
$$\barx_{\albar}=\left( x_{\al} \right)^*,\quad
 \bart_{\albar}=\left( t_{\al} \right)^*,\quad
\bary^{\albar}=\left( \my^{\al} \right)^*,\quad
\barf^{\albar}=\left( \mf^{\al} \right)^* $$
We also recall  the definition of  $h^i_A=(h^i_{\a}, h^i_{2k_- +\a}),\quad \a=1.\ldots,2k_-$
in terms of torsion and complex structures
$h^i_F=J^{i ~ AB} H_{ABF}.$

Note that equations (\ref{closedxx})(\ref{closedxxxii}) state that 
the form $C^{(s,\bars)}_{(p_1,p_2)}$ for $s=1,2$ or $\bars=1,2$ is the highest weight state
under the action of $su(2)_+\oplus su(2)_-$ on $(*,p_2)$ or $(p_1,*)$  forms on $X$. Analogously, conditions 
 (\ref{closedxix})(\ref{closedxxxvi}) imply that 
the form $C^{(s,\bars)}_{(p_1,p_2)}$ for $s=3,4$ or $\bars=3,4$ is the lowest weight state.
The corresponding $su(2)_-$ weights $iA^{-;3}=m^{(s)}_{-},\quad i{\overline A}^{-;3}=
\barm^{(\bars)}_{-}$ specify the degrees (\ref{closeddegrees}) of the form.
The following equations ensure that   
 the $su(2)_+$ weights are
$iA^{+;3}=m^{(s)}_{\pm},\quad i{\overline A}^{+;3}=\barm^{(\bars)}_{\pm}$ 
and the $u(1)$ eigenvalues are $iU=u,\, i{\overline U}=\baru.$
$$
\barf^{\delbar}\Biggl( V_{\delbar}\bigl( C^{(s,\bars)}_{\al_1 \ldots \al_{p_1}\, \belbar_1 \ldots \belbar_{p_2} }\bigr)+p_1 
{\cal F}_{\delbar \, \albar [\al_1}C^{(s,\bars)\albar}_{~~~~~~~\al_2 \ldots \al_{p_1}]\, \belbar_1 \ldots \belbar_{p_2} }\Biggr)
=-ik_-\bigl( k_-+2(\epsilon^{(s)}-iu)\bigr) C^{(s,\bars)}_{\al_1 \ldots \al_{p_1}\, \belbar_1 \ldots \belbar_{p_2} }
$$
$$
f^{\del} \Biggl ( V_{\del}\bigl( C^{(s,\bars)}_{\al_1 \ldots \al_{p_1}\, \belbar_1 \ldots \belbar_{p_2} }\bigr)-p_1 
 {\cal F}^{~~~~\bel }_{\del [\al_1}C^{(s,\bars)}_{|\bel|\al_2 \ldots \al_{p_1}]\, \belbar_1 \ldots \belbar_{p_2} }\Biggr)=
ik_-\bigl( k_-+2(\epsilon^{(s)}+iu)\bigr) C^{(s,\bars)}_{\al_1 \ldots \al_{p_1}\, \belbar_1 \ldots \belbar_{p_2} }
$$
$$
f^{\del}\Biggl( W_{\del}\bigl( C^{(s,\bars)}_{\belbar_1 \ldots \belbar_{p_2}\,
 \al_1 \ldots \al_{p_1} }\bigr)+p_2 
{\cal F}_{\del\al [\belbar_1}C^{(s,\bars)\al}_{~~~~~~~\belbar_2 \ldots \belbar_{p_2}]\, \al_1 \ldots \al_{p_1} }\Biggr)
=ik_-\bigl(- k_-+2({\overline \epsilon}^{(\bars)}+i\baru)\bigr)  C^{(s,\bars)}_{\belbar_1 \ldots \belbar_{p_2}\, \al_1 \ldots \al_{p_1} }
$$
\be
\barf^{\delbar} \Biggl ( W_{\delbar}\bigl( C^{(s,\bars)}_{\belbar_1 \ldots \belbar_{p_2}\,
 \al_1 \ldots \al_{p_1} }\bigr)-p_2 
 {\cal F}^{~~~~\albar }_{\delbar [\belbar_1}C^{(s,\bars)}_{|\albar|\belbar_2 \ldots \belbar_{p_2}]\, \al_1 \ldots \al_{p_1} }\Biggr)=
ik_-\bigl( k_-- 2({\overline \epsilon}^{(\bars)}-i\baru)\bigr) C^{(s,\bars)}_{\belbar_1 \ldots \belbar_{p_2}\, \al_1 \ldots \al_{p_1} }
\label{closedxxxxxii}
\ee
where $\epsilon^{(s)}=m_+^{(s)}+m_-^{(s)},\quad {\overline \epsilon}^{(\bars)}=
\barm_+^{(\bars)}+\barm_-^{(\bars)}.$

\subsection{The geometric meaning of the leading terms in the index}
In Section 5.2 we found the geometric interpretation of  the RR states $\vert s,\bars >$
whose orbits under spectral flow (\ref{flowii})(\ref{flowiii}) generate the contribution of $r \otimes {\overline r}$
module to
the index $I_2$. 
These states were described as $(p_1,p_2)$ forms on the target space
specified by the system of equations (\ref{closediii})-(\ref{closedxxxxxii}).

Now we clarify  the geometric meaning of the leading terms in the index $I_2.$
Ignoring  the  contribution of the excited states in
 each of the $r \otimes {\overline r}$ modules, 
we recast the expression (\ref{itwo}) for the index as 
\be
I_2=\sum_u\sum_{\baru} q^{\frac{u^2}{k}}{\overline q}^{\frac{{\overline u}^2}{k} }
\sum_{\mu=1}^{k-1}\sum_{{\overline \mu}=1}^{k-1}
q^{\frac{\mu^2}{4k^2}}\, {\overline q}^{\frac{{\overline \mu}^2}{4k^2}}\Biggl(
z^{\mu}\, {\barz}^{~\overline {\mu}}d_{++}(\mu,{\overline \mu})+
z^{\mu}\, {\barz}^{~-\overline {\mu}}d_{+-}(\mu,{\overline \mu})+
\label{indexgeom}
\ee
$$
z^{-\mu}\, {\barz}^{~\overline {\mu}}d_{-+}(\mu, {\overline \mu})+
z^{-\mu}\, {\barz}^{~-\overline {\mu}}d_{--}(\mu, {\overline \mu})\Biggr)+\ldots
$$
In (\ref{indexgeom}) we denote:
\be
d_{++}(\mu, {\overline \mu})=\sum_{p_1 \ = \ max( \mu+k_- - k_+, \ k_-)}^{min(k_- +\mu-1,\ 2k_- -1)}\, \sum_{p_2 \ = \ max(k_- - \overline{\mu} +1, \ 1)}^{min(k-{\overline \mu},\ k_-)}
(-)^{p_1+p_2}\, n(p_1,+;p_2,+)
\label{dofmu}
\ee

\be
d_{+-}(\mu, {\overline \mu})=-\sum_{p_1 \ = \ max( \mu+k_- - k_+, \ k_-)}^{min(k_- +\mu-1,\ 2k_- -1)}\,\sum_{p_2 \ = \ max( {\overline \mu} + k_- - k_+, \ k_-)}^{min(k_- +{\overline \mu}-1,\ 2k_- -1)}
(-)^{p_1+p_2}\, n(p_1,+;p_2,-)
\label{dofmuiii}
\ee

\be
d_{-+}(\mu, {\overline \mu})=-\sum_{p_1 \ = \ max(k_- - \mu+1, \ 1)}^{min(k-\mu,\ k_-)} \,
\sum_{p_2 \ = \ max(k_- - \overline{\mu} +1, \ 1)}^{min(k-{\overline \mu},\ k_-)}
(-)^{p_1+p_2}\, n(p_1,-;p_2,+)
\label{dofmuiv}
\ee

\be
d_{--}(\mu,{\overline \mu})=\sum_{p_1 \ = \ max(k_- - \mu+1, \ 1)}^{min(k-\mu,\ k_-)}\,
 \sum_{p_2 \ = \ max( {\overline \mu} + k_- - k_+, \ k_-)}^{min(k_- +{\overline \mu}-1,\ 2k_- -1)}
(-)^{p_1+p_2}\, n(p_1,-;p_2,-)
\label{dofmuii}
\ee 
Here $n(p_1,+;p_2,+)$ is the number of  $(p_1,p_2)$ forms which
 have the following special properties:
\begin{enumerate}
\item
\be
{\cal D} C= \nu \wedge C,
\quad  {\cal D}^{\dg} C=0,\quad 
{\barcalD} C= 0,\quad 
\quad {\barcalD}^{\dg} C=i_{\displaystyle{\bar w}}C 
\label{conclusioni}
\ee
where (anti-)holomorphic differential $\left({\barcalD}\right){\cal D}$ with
torsion is defined in (\ref{differ}). 
The symbol $i_{\displaystyle {\bar w}}$ denotes the interior product with vector ${\bar w}.$
The 1-form $\nu$ and vector
$\barmw$ have flat components
\be
\nu_{\al}=\frac{i}{2k_-}
t_{\al}\bigl(\mu +i u\bigr),
\quad 
\barmw^{\albar}=\frac{i}{2k_-}
\barf^{\albar}\bigl(\barmu +i \baru\bigr)
\label{conclusioniii}
\ee
where $\barf^{\albar}=t_{\al}=h^3_{\a}+ih^3_{2k_-+\a},\quad 
\a=1,\ldots, 2k_-$
and we recall that $h^i_A=(h^i_{\a},h^i_{2k_-+\a})$  
is defined as $h^i_A=J^{i~BC}H_{ABC}.$
\item
\be
C= *_{(J^-,\, J^+)} B,\quad {\cal D}^{\dg}B=0,\quad {\barcalD}B=0
\label{conclusioniv}
\ee
where Hodge-like duality operation $*_{(J^-,\, J^+)}$ is defined
in (\ref{closedvii}).
\item 
$$
J^+_{[ \al \bel} C_{\al_1 \ldots \al_{p_1}]\, \belbar_1 \ldots \belbar_{p_2} }=0,
\quad
\bary^{\delbar} V_{\delbar}\bigl( C_{\al_1 \ldots \al_{p_1}\, \belbar_1 \ldots \belbar_{p_2} }\bigr)=0,$$
\be
J^{+ \  \albar \belbar} C_{\albar \belbar \, \belbar_3 \ldots \belbar_{p_2} \, \al_1 \ldots \al_{p_1}   }=0, \quad
\bary^{\delbar}W_{\delbar}\bigl( C_{\belbar_1 \ldots \belbar_{p_2}\, \al_1 \ldots \al_{p_1}  }\bigr)=0
\label{conclusionv}
\ee
where vector fields $V_{\al},V_{\albar},\ W_{\al},W_{\albar}$
are given in (\ref{eigen},\ \ref{eigenclosedii}) and $\bary^{\albar}=h^1_{\a}+ih^2_{\a}.$
\item
\be
x \wedge C=0,\quad i_{\barmy}C=0,\quad \bart \wedge C=0,\quad i_{\displaystyle \! f}C=0
\label{conclusionvi}
\ee
where 1-forms $x,\bart$ and vectors $\barmy,\mf$ have flat components
$$x_{\al}=\barmy^{\albar}, \quad \bart_{\albar}=\mf^{\al}=h^3_{\a}-ih^3_{2k_-+\a}$$
\end{enumerate}

The equations in group III state that the
form counted in $n(p_1,+;p_2,+)$ has
the highest weight under the action of
$su(2)_+\oplus su(2)_-$ algebra on $(*,p_2)$ and $(p_1,*)$ forms on $X.$ 
The corresponding weights are:
$$  iA_0^{-;3}=\frac{p_1-k_-}{2},\quad i{\overline A}_0^{-;3}=\frac{k_--p_2}{2},\quad
iA_0^{+;3}=\frac{\mu-p_1+k_-}{2},\quad  
 i{\overline A}_0^{+;3}=\frac{{\overline \mu}- k_- + p_2}{2} $$
The eigenvalues of the form  under the $u(1)$ action are
$$ iU=u,\quad i\overline U=\baru$$ 
These eigenvalue equations have the structure (\ref{closedxxxxxii})
with substitution $\epsilon^{(s)} \to \mu, \quad {\overline \epsilon}^{(\bars)} \to \barmu$

Analogously, $n(p_1,+;p_2,-)$ is the number of  $(p_1,p_2)$ forms 
which
 have the following special properties:
\begin{enumerate}
\item 
\be
{\cal D} C= \nu' \wedge C,
\quad  {\cal D}^{\dg} C=0,\quad 
{\barcalD} C= \barnu' \wedge C,\quad 
\quad {\barcalD}^{\dg} C=0 
\label{newconclusioni}
\ee
where 
\be
\nu'_{\al}=\frac{i}{2k_-}
t_{\al}\bigl(\mu +i u\bigr),
\quad 
\barnu'_{\albar}=-\frac{i}{2k_-}f^{\al}
\bigl(-\barmu -i \baru\bigr),
\label{newconclusionii}
\ee
\item 
\be
C= *_{(J^-,\, J^+)} B',\quad {\cal D}^{\dg}B'=0,\quad {\barcalD}^{\dg}B'=0
\label{newconclusioniv}
\ee
\item 
$$
J^+_{[ \al \bel} C_{\al_1 \ldots \al_{p_1}]\, \belbar_1 \ldots \belbar_{p_2} }=0,
\quad
\bary^{\delbar} V_{\delbar}\bigl( C_{\al_1 \ldots \al_{p_1}\, \belbar_1 \ldots \belbar_{p_2} }\bigr)=0,$$
\be
J^{-}_{[ \albar \belbar} C_{ \belbar_1 \ldots \belbar_{p_2}] \, \al_1 \ldots \al_{p_1}   }=0, \quad
y^{\del}W_{\del}\bigl( C_{\belbar_1 \ldots \belbar_{p_2}\, \al_1 \ldots \al_{p_1}  }\bigr)=0
\label{newconclusionv}
\ee
\item 
\be
x \wedge C=0,\quad \quad \barx \wedge C=0, \quad i_{\barmf}C=0,\quad i_{\mf}C=0
\label{newconclusionvi}
\ee
\end{enumerate}
The equations in group III state that the
form counted in $n(p_1,+;p_2,-)$ has
the highest weight under the action of
$su(2)_+\oplus su(2)_-$ algebra on $(*,p_2)$ forms and the lowest weight under
the action on $(p_1,*)$ forms. 
The corresponding weights are:
$$  iA_0^{-;3}=\frac{p_1-k_-}{2},\quad i{\overline A}_0^{-;3}=\frac{k_--p_2}{2},\quad
iA_0^{+;3}=\frac{\mu-p_1+k_-}{2},\quad  
 i{\overline A}_0^{+;3}=\frac{p_2-{\overline \mu}- k_- }{2} $$
The eigenvalues of the form  under the $u(1)$ action are
$$ iU=u,\quad i\overline U=\baru$$ 
These eigenvalue equations have the structure (\ref{closedxxxxxii})
with substitution $\epsilon^{(s)} \to \mu,\quad {\overline \epsilon}^{(\bars)} \to - \barmu$

The other two quantities $n(p_1,-;p_2,+)$ and $n(p_1,-;p_2,-)$ can be defined
in a similar way.

\section{Conclusion}
The main result of this paper is a geometric interpretation  of the index $I_2$
for the $N=4$ gauged WZW models.  
In particular, we showed that  the states contributing to the index $I_2$ 
belong to spectral flow orbits (\ref{flowii})(\ref{flowiii})  of special RR ground states. We 
characterized these states  geometrically as  $(p_1,p_2)$ forms
on the target space specified by equations (\ref{closediii})-(\ref{closedxxxxxii}).

Besides, we showed that the coefficients of the various leading terms in 
 the index  can be obtained by   
 counting with $(-1)^{p_1+p_2}$ sign the numbers $n(p_1,p_2)$ of $(p_1, p_2)$ 
forms with certain properties (\ref{conclusioni})-(\ref{newconclusionvi}).
This is similar in spirit to  the  geometric description of the
leading contribution  to the elliptic genus but the forms counted in the
index $I_2$
are more special. In particular, they have either the highest or the lowest
weight  under the action of $su(2)_+\oplus su(2)_-$ algebra on
 differential forms on the target space.

Although we have used the structure of the $N=4$ cosets in realization of $A_{\c}$
algebra,  our results are formulated in a rather general form in terms of the torsion 
and the triplet of complex structures  on the target space. 
Therefore, it is a natural question  if our geometric interpretation is valid
for arbitrary $\sigma$-models with $A_{\c}$ symmetry.
It is hard to answer  this question. The major problem stems from  
the fact that  realization of $A_{\c}$ algebra is known only
for $N=4$ gauged WZW models. One can  construct new theories 
with   $A_{\c}$ symmetry by considering symmetric product orbifolds of these WZW
models, but then one has to deal with another problem of smoothing the target space
into a manifold by 
resolving the orbifold singularities.
Finding a general $\sigma$-model with $A_{\c}$ symmetry 
remains an open problem.

{\bf Acknowledgements}

I would like to thank G. Moore and S. Gukov  for  valuable discussions. 
I am also grateful to  A. Adams, L. Motl and  E. Yuzbashyan
for the comments about this paper.  
My research was supported in part
by NSF grants PHY-0244821 and DMS-0244464. 

\section{Appendix A}
Here we give the component form of the action for the supersymmetric gauged WZW model.
As we review in section 2 the path integral for this theory
takes the form:
\be
{\bf Z}=\int [d\bG][d\bH][d\bB][d\bC][d\barbB][d\barbC]e^{-\int d^2\tht \Bigl( I(\bG)-I(\bH)\Bigr )-S_{ghost}} 
\label{app}
\ee
where the $\sigma$-model fields are arranged  in superfields as:
\be
\bG=g\Bigl( 1+\tht \psi +\btht g^{-1} \bpsi g - \tht \btht \psi g^{-1}\bpsi g \Bigr)
\label{appii}
\ee
\be
\bH=h\Bigl( 1+\tht \eta +\btht h^{-1} {\overline \eta} h - \tht \btht \eta h^{-1} {\overline \eta}h \Bigr)
\label{appiii}
\ee
where $g(h)$ is a map from Riemann surface $\Sigma$ to the group  $G(H).$
The fermions $\psi,\ \bpsi (\eta \ ,\overline{\eta} )$ take values in the Lie algebra of G(H).

The component form of the action is given by: 
\be
S=I_{bos}(g)-I_{bos}(h)+I_{ferm}(\psi,\bpsi,g)-I_{ferm}(\eta,\overline{\eta},h)+S_{ghost}
\label{appcomp}
\ee
where
\be
I_{bos}(g)=-\frac{k}{8 \pi}\Biggl \{
\int_{\Sigma} d^2z Tr'\Biggl( g^{-1} \p g g^{-1} {\overline {\p}} g \Biggr)
+ \int_{B} dt d^2z Tr' \Biggl({\tilde g}^{-1} \p_t {\tilde g} 
\Bigl[{\tilde g}^{-1} \p \tilde g, {\tilde g}^{-1} {\overline {\p}} {\tilde g}\Bigr] \Biggr) \Biggr\}
\label{appv}
\ee
and
\be
I_{ferm}(\psi,\bpsi,g)=\frac{k}{8 \pi}
\int_{\Sigma} d^2z \Biggl \{Tr'\Bigl(\psi {\overline \nabla }_g\psi \Bigr)+Tr'\Bigl(\bpsi 
{ \nabla }_g\bpsi \Bigr)\Biggr \}
\label{appferm}
\ee
Here $\nabla_g=\p-[\p gg^{-1},-],\quad  {\overline \nabla}_g={\overline \p}-[g^{-1}{\overline \p} g
,-]$
are the covariant derivatives in the adjoint
representation of the Lie algebra of $G$.

Introducing coordinates $X^{\hM}$ on $G$ we define a veilbein:
\be
E_{\hM}^{\hA}=-i \sqrt{\frac{k}{2}}\bigl( \p_{\hM}g g^{-1}\Bigr)^{\hA}, \quad 
E_{\hM}^{\hA}E^{\hN}_{\hA}=\delta_{\hM}^{\hN}
\label{veil}
\ee 
so that $I_{bos}(g)$ has a standard $\sigma$-model form (with $\alpha'=1$) 
\be
I_{bos}(g)=\frac{1}{2 \pi} \int_{\Sigma} d^2z 
\Bigl( g_{\hM \hN}+B_{\hM \hN}\Bigr)\p X^{\hM}  {\overline {\p}}  X^{\hN}  
\label{appvii}
\ee
Here $g_{\hM \hN}=E_{\hM}^{\hA}E_{\hN}^{\hB}\delta_{\hA \hB}$
and $B_{\hM \hN}$ is defined as
\be
\frac{1}{\sqrt{k}}E_{\hM}^{\hA}E_{\hN}^{\hB}E_{\hP}^{\hC}H_{\hA \hB \hC}
=-\frac{3}{2}\p_{[\hM} B_{\hN \hP]}
\label{appbf}
\ee
\section{Appendix B}
Here we summarize the OPE's of the $A_{\c}$ algebra:
\be
G^{a}(z)G^{b}(w)\sim \frac{2c}{3} \frac{\delta^{ab} }{(z-w)^3}
-\frac{4k_- t^{+i}_{ab}A^{+i}(w)+ 4k_+ t^{-i}_{ab}A^{-i}(w)}{k(z-w)^2}
\label{appbi}
\ee
$$-\frac{2k_- t^{+i}_{ab}\p A^{+i}(w)+ 2k_+ t^{-i}_{ab}\p A^{-i}(w)}{k(z-w)}
+\frac{2\delta^{ab} T(w)}{z-w} $$

\be
A^{i\pm}(z)A^{j\pm}(w)\sim -\frac{k_{\pm}}{2}\frac{\delta^{ij}}{(z-w)^2}+
\frac{\varepsilon_{ijk}A^{k\pm}}{z-w}
\label{appbiii}
\ee

\be
Q^{a}(z)G^{b}(w)\sim  \frac{\delta^{ab} U(w) }{(z-w)}
+\frac{t^{+i}_{ab}A^{+i}(w)-t^{-i}_{ab}A^{-i}(w)}{(z-w)}
\label{appbiv}
\ee
 
\be
A^{\pm;i}(z)G^{a}(w)\sim \mp \frac{k_{\pm}}{k}\frac{t^{\pm i}_{ab} Q^b(w) }{(z-w)^2}
+\half \frac{t^{\pm i}_{ab}G^b(w)}{(z-w)}
\label{appbv}
\ee

\be
A^{\pm;i}(z)Q^{a}(w)\sim \half \frac{t^{\pm i}_{ab}Q^b(w)}{(z-w)}
\label{appbvi}
\ee
  \be
Q^{a}(z)Q^{b}(w)\sim - \frac{k}{2} \frac{\delta^{ab}  }{(z-w)}
\label{appbvii}
\ee
\be
U(z)G^{a}(w)\sim  \frac{Q^a(w) }{(z-w)^2}
\label{appbviii}
\ee
 \be
U(z)U(w)\sim - \frac{k}{2} \frac{1  }{(z-w)^2}
\label{appbix}
\ee
where
$t^{\pm i}_{ab}=\pm 2\delta^i_{[a}\delta^4_{b]}+ \e_{iab}$ and 
$$
c=\frac{6k_{+}k_{-}}{k},\quad k=k_{+}+k_{-}
\label{appbii}
$$
Comment: The name $A_{\c}$ originated from the fact that the OPE's of the algebra
can be parametrized by two parameters $k$ and $\c=\frac{k_-}{k}.$
\section{Appendix C}
Here we give explicit expressions for $H_{ABC}, J^i_{AB}, h^a_F, M^i_{AB}$
in the example  of $SU(3)$ WZW model.
The hermitean generators of $su(3)$ are:
\be
T^1=\frac{1}{\sqrt{2}}\left(\begin{array}{ccc}
0 & -i & 0 \\ 
i & 0 & 0 \\
0 & 0 & 0 \\ \end{array}\right),
\quad \quad
T^2=\frac{1}{\sqrt{2}} \left(\begin{array}{ccc}
0 & 1 & 0 \\ 
1 & 0 & 0 \\
0 & 0 & 0 \\ \end{array}\right)
\label{appendixci}
\ee
\be
T^3=\frac{1}{\sqrt{2}}\left(\begin{array}{ccc}
0 & 0 & 0 \\ 
0 & 0 & -i \\
0 & i & 0 \\ \end{array}\right), \quad \quad
T^4= \frac{1}{\sqrt{2}}\left(\begin{array}{ccc}
0 & 0 & 0 \\ 
0 & 0 & 1 \\
0 & 1 & 0 \\ \end{array}\right)
\label{appendixcii}
\ee
\be
T^5=\frac{1}{\sqrt{2}}\left(\begin{array}{ccc}
0 & 0 & -i \\
 0 & 0& 0\\
i & 0 & 0\\ \end{array}\right), \quad \quad
T^6= \frac{1}{\sqrt{2}}\left(\begin{array}{ccc}
0 & 0 & 1 \\
 0 & 0& 0\\
1 & 0 & 0\\ \end{array}\right)
\label{appendixciii}
\ee
\be
T^7=\frac{1}{\sqrt{2}}\left(\begin{array}{ccc}
 1 & 0 & 0 \\
0 & 0& 0\\
0 & 0 & -1 \\ \end{array}\right), 
\quad  \quad 
T^8=\frac{1}{\sqrt{6}}\left(\begin{array}{ccc}
 1 & 0 & 0 \\
0 & -2 & 0\\
0 & 0 & 1 \\ \end{array}\right)
\label{appendixciv}
\ee
They satisfy $[T^A,T^B]=if_{ABC}T^C, \quad Tr T^A T^B=\delta^{AB}.$

The non-zero components of torsion $H_{ABC}=\frac{1}{\sqrt{2}}f_{ABC}$:
\be
H_{127}=-\half, \quad H_{128}=-\frac{\sqrt{3}}{2},\quad
H_{347}=-\half, \quad H_{348}=\frac{\sqrt{3}}{2}, 
\label{appendixcv}
\ee
\be
H_{135}=-\half, \quad H_{146}=-\frac{1}{2},\quad
H_{245}=\half, \quad H_{236}=-\frac{1}{2},\quad H_{567}=-1 
\label{appendixcvi}
\ee
The non-zero components of complex structures $J^i_{AB}$
\be
J^3_{12}=-1, \quad J^3_{34}=-1,\quad J^3_{56}=-1, \quad J^3_{78}=-1
\label{appendixcvii}
\ee
\be
J^2_{13}=-1, \quad J^2_{24}=1,\quad J^2_{58}=-1, \quad J^2_{67}=-1
\label{appendixcviii}
\ee
\be
J^1_{14}=1, \quad J^1_{23}=1,\quad J^1_{57}=-1, \quad J^1_{68}=1
\label{appendixcix}
\ee
The non-zero components of $h^a_F$ defined in (\ref{hbf}):
\be
h^1_F=-4 \delta_{F,6},\quad h^2_F=4 \delta_{F,5},\quad
h^3_F=4\delta_{F,7},\quad h^4_F=-4 \delta_{F,8}
\label{appendixcx}
\ee
The non-zero components of $M^i_{AB}$ defined in (\ref{hif}):
\be
M^3_{78}=-2,\quad M^3_{56}=2,\quad M^1_{57}=2,\quad M^1_{68}=2
\label{appendixcxi}
\ee
\be
M^2_{58}=-2,\quad M^2_{67}=2
\label{appendixcxii}
\ee
\section{Appendix D}
Here we give  explicit expressions for the characters
which appear in Section 3.

The  $ A_{\gamma}$ character of the model $S$ 
is given by:
\be
 {\cal S}^R\bigl(u;q,z_{\pm}\bigr)=q^{u^2/k+1/8}F^R(q,z_{\pm})\times
\prod_{n=1}^{\infty} (1-q^n)^{-1}\bigl(z_{+}+z_{-}+z_{+}^{-1}+z_{-}^{-1}\bigr)
\ee
where 
\be
F^R(q,z_{\pm})=\prod_{n=1}^{\infty}(1+z_{+}z_{-}q^n)(1+z_{+}^{-1}z_{-}q^n)(1+z_{+}z_{-}^{-1}q^n)
(1+z_{+}^{-1}z_{-}^{-1}q^n)
\label{fr}
\ee

The character of 
the massless R-sector representation  ${\tilde r}=(\l_+,\l_-)$ of  the $\A_{\gamma}$
algebra has the following form:
\be
Ch_0^{{\tilde A}_{\gamma},\  R} 
\bigl( {\tilde l}_+,\l_-; q,z_{\pm}\bigr)=
q^{{\tilde h}-{\tilde c}/24}F^R(q,z_{\pm})\times
\label{tildech}
\ee
$$\prod_{n=1}^{\infty}(1-q^n)^{-2}(1-z_{+}^2q^n)^{-1}
(1-z_{+}^{-2}q^n)^{-1}(1-z_{-}^{2}q^n)^{-1}
(1-z_{-}^{-2}q^n)^{-1}$$
$$\prod_{n=1}^{\infty}(1-q^n)^{-1}(z_{+}^{-1}+z_{-}^{-1})(1+z_{+}^{-1}z_{-}^{-1})
(1-z_{+}^{-2})^{-1}(1-z_{-}^{2})^{-1}$$
$$\times \sum_{m,n=-\infty}^{\infty}q^{n^2k_{+}+2n\l_{+}+
m^2k_{-}+2m\l_{-}} $$
$$ \sum_{\e_{+},\e_{-}=\pm 1} \e_{+}\e_{-}z_{+}^{2\e_{+}(\l_{+}+nk_{+})}
z_{-}^{2\e_{-}(\l_{-}+mk_{-})}\bigl( z_{+}^{-\e_{+}} q^{-n} + z_{-}^{-\e_{-}} q^{-m}\bigr )^{-1}
$$
where
\be
{\tilde h}=\frac{ (k_+ -1 )(k_- -1 )}{4k}+
\frac{ (\l_{+}+\l_{-})(\l_{+}+\l_{-}+1)}{k},
\label{tildehw}
\ee
$$
{\tilde c}=\frac{6k_{+}k_{-}}{k}-3,\quad k=k_{+}+k_{-}.$$

\end{document}